\documentclass[rmp,amssymb,showpacs,showkeys,preprint,named]{revtex4}
\bibpunct{[}{]}{,}{a}{}{;}
\usepackage{graphicx}
\usepackage{eepic}
\RequirePackage{times}
\RequirePackage{mathptm}
\usepackage{xcolor}
\begin{document}

\title{Contexts in quantum, classical and partition logic}

\author{Karl Svozil}
\email{svozil@tuwien.ac.at}
\homepage{http://tph.tuwien.ac.at/~svozil}
\affiliation{Institute for Theoretical Physics, Vienna University of Technology,
Wiedner Hauptstra\ss e 8-10/136, A-1040 Vienna, Austria}

\begin{abstract}
Contexts are maximal collections of co-measurable observables ``bundled together'' to form a ``quasi-classical mini-universe.'' Different notions of contexts are discussed for classical, quantum and generalized urn--automaton systems.
\end{abstract}


\maketitle

%
%
%
%
%
%
%
%
%
%
%
%
%

\tableofcontents

\newpage

{\footnotesize
\begin{quote}
\begin{flushright}
{
It is not enough
to have no concept,                \\
one must also be capable
of expressing it.\footnote{Es gen{\"{u}}gt nicht, keinen Gedanken zu haben:
man mu\ss ~ ihn auch ausdr{\"{u}}cken k{\"{o}}nnen.}\\
{\em Karl Kraus}, in {\it Die~Fackel~697,~60~(1925)} } \\
$\;$\\
But no sooner do we depart from sense and instinct to follow the light of a superior principle, to reason, meditate, and reflect on the nature of things, but a thousand scruples spring up in our minds concerning those things which before we seemed fully to comprehend. Prejudices and errors of sense do from all parts discover themselves to our view; and, endeavouring to correct these by reason, we are insensibly drawn into uncouth paradoxes, difficulties, and inconsistencies, which multiply and grow upon us as we advance in speculation, till at length, having wandered through many intricate mazes, we find ourselves just where we were, or, which is worse, sit down in a forlorn Scepticism
\\
{\em George Berkeley}, in {\it A Treatise Concerning the Principles of Human Knowledge (1710)}
\end{flushright}
\end{quote}

}


\section{Motivation}

In what follows, the term
\index{context}
{\em context} refers to a maximal collection of co-measurable observables ``bundled together'' to form  a ``quasi-classical mini-universe''
within some ``larger'' nonclassical structure.
Similarly, the contexts of an observable are often defined as maximal collections of mutually co-measurable (compatible) observables
which are measured or at least could in principle be measured alongside of this observable \cite{bohr-1949,bell-66,hey-red,redhead}.
Quantum mechanically, this amounts to a formalization of contexts by Boolean subalgebras of Hilbert lattices \cite{svozil-2004-analog,svozil-2004-vax},
or equivalently, to maximal operators
(e.g., Ref.~\cite[Sec.~II.10, p. 90]{v-neumann-49},
English translation in Ref.~\cite[p.~173]{v-neumann-55}, Ref.~\cite[\S~2]{kochen1}, Ref.~\cite[pp.~227,228]{neumark-54}, and  Ref.~\cite[\S~84]{halmos-vs}).

In classical physics, contexts are rather unrevealing, as all classical observables are in principle co-measurable,
and there is only a single context which comprises the entirety of observables.
Indeed, that two or more observables may not be co-measurable; i.e., operationally obtainable
simultaneously, and thus may belong to different, distinct contexts, did not bother the classical mind until around 1920.
This situation has changed dramatically with the emergence of quantum mechanics, and in particular
with the discovery of complementarity and value indefiniteness.
Contexts are the building blocks of quantum logics;
i.e., the pastings of a continuity of contexts form the Hilbert lattices.

We shall make use of algebraic formalizations, in particular logic.
Quantum logic is about the relations and operations among statements
referring to the quantum world.
As quantum physics is an extension of classical physics,
so is quantum logic an extension of classical logic.
Classical physics can be extended in many mindboggling, weird ways.
The question as to why Nature ``prefers'' the quantum mindboggling way over others
appears most fascinating to the open mind.
Before understanding some of the issues,
one has to review classical as well as quantum logic and some of its doubles.

Logic will be expressed as algebra.
That is an approach which can be formalized.
Other approaches, such as the widely held opportunistic belief
that something is true because it is useful
might also be applicable (for instance in acrimonious divorces),
though less formalized.
Some of the material presented here has already been published elsewhere
\cite{svozil-ql},
in particular the partition logic part \cite{svozil-2001-eua}, or
the section on quantum probabilities
\cite{svozil-tkadlec}. Here we emphasize the importance of the notion of {\em context},
which may serve as a unifying principle for all of the logics discussed.

\section{Classical contexts}

Logic is an ancient philosophical discipline.
Its algebraization started in the mid-nineteenth century with
Boole's {\em Laws of Thought} \cite{Boole}.
In what follows, Boole's approach, in particular to probability theory, is reviewed.

\subsection{Boolean algebra}
\index{Boolean algebra}

A Boolean algebra ${\mathfrak B}$ is a set endowed with
two binary operations $\wedge$ (called ``and'')  and  $\vee$ (called ``or''),
as well as a unary operation ``~$'$~'' (called ``complement'' or ``negation'').
It also contains two elements $1$ (called ``true'') and $0$ (called ``false'') satisfying
associativity,   commutativity, the  absorption law and   distributivity.
Every element has a unique complement.

A typical example of a Boolean algebra is set theory. The operations
are identified with the set theoretic intersection, union, and complement, respectively.
The implication relation is identified with the subset relation.

\subsection{Classical contexts as classical logics}

A classical Boolean algebra is the representation
of all possible ``propositions'' or ``knowables.''
Every knowable can be combined with every other one by the standard logical operations
``and'' and ``or.''
Operationally,
all knowables are in principle knowable simultaneously.
Stated differently:
within the Boolean ``universe,'' the knowables are
all consistently co-knowable.
In this sense, classical contexts coincide with the collection of all possible observables,
which are expressed by Boolean algebras.
Thus, classical contexts can be identified with the respective classical logics.

\subsection{Classical probabilities}

Classical probabilities and joint probabilities can be represented as
points of a {\em convex polytope}
\index{convex polytope}
spanned by all possible ``extreme cases'' of the classical Boolean algebra; more formally:
by all two-valued measures on the Boolean algebra. Two-valued measures, also
called dispersionless measures or valuations, acquire only the values ``0'' and ``1,''
interpretable as  falsity and truth, respectively.
If some events are independently measured, then their joint probability $pq\cdots$ can be expressed as the product of their individual probabilities
$p$, $q$, $\ldots$.

The associated {\em correlation polytope}
\cite{pitowsky,pitowsky-89a,Pit-91,Pit-94,2000-poly}
(see also Refs.~\cite{froissart-81,cirelson:80,cirelson})
is  spanned by a convex combination  of vertices, which are vectors of the form $(p,q, \ldots, pq,\ldots )$,
where the components are the individual probabilities of independent events which take on the values $0$ and $1$,
together with their joint probabilities, which are the products of the individual probabilities.
The polytope faces impose  ``inside--outside'' distinctions.
The associated inequalities must be obeyed
by all classical probability distributions; they are  bounds on classical (joint) probabilities
termed {\em ``conditions of possible experience''} by Boole \cite{Boole,Boole-62}.

\subsubsection{Two-event ``1--1'' case}
\begin{figure}
  \centering
\includegraphics[width=90mm]{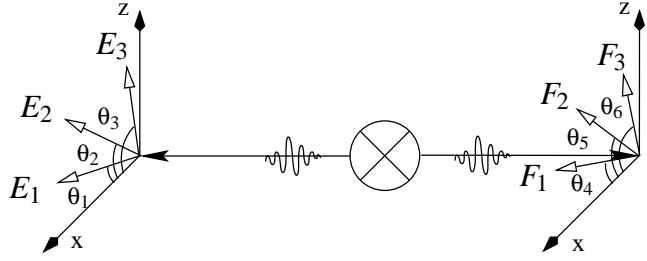}
  \caption{Measurements of  $E_1,E_2,E_3$ on the ``left,'' and $F_1,F_2,F_3$ on the ``right'' hand side, along directions $\theta_i$.
  \label{2006-ql-2} }
\end{figure}
Let us demonstrate the bounds on classical probabilities by the simplest nontrivial example of two propositions;
e.g.,
\begin{quote}
$E\equiv${\em ``a particle detector aligned along direction ${\bf a}$ clicks,''}
and\\
$F\equiv${\em ``a particle detector aligned along direction ${\bf b}$ clicks.''}
\end{quote}
Consider also the joint proposition
\begin{quote}
$E\wedge F \equiv$ {\em ``the two particle detectors
aligned along directions ${\bf a}$ and ${\bf b}$ click.''}
\end{quote}
The notation ``1--1'' alludes to the experimental setup, in which
the two events are registered by detectors located at two ``adjacent sites.''
For multiple direction measurements, see Fig.~\ref{2006-ql-2}.

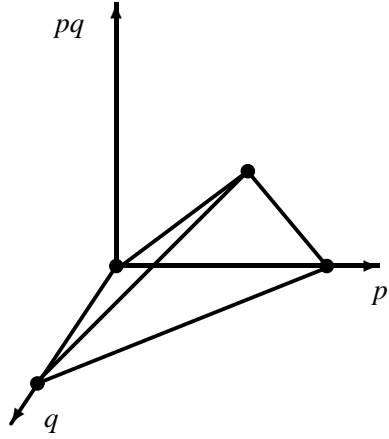
\begin{figure}
\centerline{
\unitlength 0.7mm
\allinethickness{1.5pt} 
\begin{picture}(70.00,80.00)
\put(20.00,30.00){\vector(0,1){50.00}}
\put(20.00,30.00){\vector(1,0){50.00}}
\put(20.00,30.00){\vector(-2,-3){20.00}}
\put(60.00,30.00){\circle*{2.00}}
\put(5.00,7.67){\circle*{2.00}}
\put(45.00,48.00){\circle*{2.00}}
\put(20.00,30.00){\circle*{2.00}}
\put(5.00,7.67){\line(5,2){55.00}}
\put(60.00,30.00){\line(-5,6){15.00}}
\put(45.00,48.00){\line(-1,-1){40.00}}
\put(45.00,48.00){\line(-4,-3){25.00}}
\put(70.00,24.33){\makebox(0,0)[cc]{$p$}}
\put(7.67,0.00){\makebox(0,0)[cc]{$q$}}
\put(11.00,75.00){\makebox(0,0)[cc]{$pq$}}
\end{picture}
}
\caption{The correlation polytope for two events.
The vertices are $(0,0,0)$,
$(0,1,0)$,
$(1,0,0)$, and
$(1,1,1)$. The four faces of the polytope are characterized by the inequalities
in Table~\ref{2006-ql-t1}(b).
 \label{2006-ql-f1}
}
\end{figure}

There exist four possible cases,
enumerated in Table  \ref{2006-ql-t1}(a).
\begin{table}
\centerline{
{\footnotesize
 \begin{tabular}{cc}
\qquad\begin{tabular}{|c|ccc|}
\hline\hline
& $\quad E\quad $ & $\quad F\quad $ & $\quad E\wedge F \equiv E \cdot F\quad $\\
\hline
$1$& 0&0&0   \\
$2$& 0&1&0   \\
$3$ & 1&0&0  \\
$4$ & 1&1&1  \\
\hline\hline
\end{tabular}
\quad
&
\quad
\begin{tabular}{|c|c|}
\hline\hline
 & full facet inequality\\
\hline
$1$& $pq \ge 0$ \\
$2$& $p \ge pq$   \\
$3$ & $p \ge pq$  \\
$4$ & $pq \ge p+q-1$  \\
\hline\hline
\end{tabular}
\qquad \\ \\
(a)&(b)\\
\end{tabular}
}
}
\caption{Construction of the correlation polytope for two  events: (a) the four possible cases are represented by
the truth table, whose rows can be interpreted as three-dimensional vectors forming the vertices of the correlation polytope;
(b) the resulting four faces of the polytope are characterized by half-spaces
which are obtained by solving the hull problem.
 \label{2006-ql-t1}.
}
\end{table}
The correlation polytope in this case is formed by interpreting the rows as vectors
in three-dimensional vector space. Four cases, interpretable as truth assignments or two-valued measures,
correspond to the four vectors
$(0,0,0)$,
$(0,1,0)$,
$(1,0,0)$, and
$(1,1,1)$.
The correlation polytope for the probabilities $p$, $q$ and the joint probabilities $pq$ of an occurrence of $E$, $F$, and both $E\& F$
$$ (p,q,pq)=
\kappa_1 (0,0,0)+
 \kappa_2 (0,1,0)+
 \kappa_3 (1,0,0)+
 \kappa_4 (1,1,1)=(\kappa_3+\kappa_4, \kappa_2+\kappa_4,\kappa_4)$$
is spanned by the convex sum $ \kappa_1 +
 \kappa_2 +
 \kappa_3 +
 \kappa_4 =1$
of these four vectors,
which thus are vertices of the polytope.
$\kappa_i$ can be interpreted as the normalized weight for event $i$ to occur.
The configuration is drawn in Figure~\ref{2006-ql-f1}.

By the
Minkoswki-Weyl representation
\index{Minkoswki-Weyl representation}
theorem (e.g, Ref.~\cite[p.29]{ziegler}),
every convex polytope has a dual (equivalent) description:
either as the convex hull of its extreme points (vertices);
or as the intersection of a finite number of half-spaces.
Such facets are given by linear inequalities,
which are obtained
from the set of vertices
by solving the so called
{\em hull problem}.
\index{hull problem}
The inequalities coincide with Boole's ``conditions of possible experience.''
The hull problem is algorithmically solvable but computationally hard \cite{pit:90}.

In the above example, the
``conditions of possible experience'' are given by the inequalities enumerated in
Table~\ref{2006-ql-t1}b).
One of their consequences are bounds on joint occurrences of events.
Suppose, for example, that
the probability of a click in detector aligned along direction ${\bf a}$ is 0.9, and
the probability of a click in the second detector aligned along direction ${\bf b}$ is 0.7.
Then inequality 4 forces us to accept that the probability
that both detector register clicks cannot be smaller than $0.9+0.7-1= 0.6$.
If, for instance, somebody comes up with a joint probability of $0.4$, we would
know that this result is flawed,
possibly by fundamental measurement errors, or by cheating, or by (quantum) ``magic.''

\subsubsection{Four-event ``2--2'' case}
A configuration  discussed in quantum mechanics is one with four events
grouped into two equal parts $E_1,E_2$ and $F_1,F_2$.
There are $2^4$ different cases  of occurrence or nonoccurrence of these four events
enumerated in Table
 \ref{2006-ql-t2}.

\begin{table}
\centerline{
{\footnotesize
\begin{tabular}{|c|cccccccc|ccccc}
\hline\hline
& $ E_1$ & $ E_2$ & $F_1$& $F_2$& $ E_1F_1$ & $ E_1F_2$ & $E_2F_1$& $E_2F_2$\\
\hline
1   &0  &0  &0  &0  &0  &0  &0  &0    \\
2   &0  &0  &0  &1  &0  &0  &0  &0    \\
3   &0  &0  &1  &0  &0  &0  &0  &0    \\
4   &0  &0  &1  &1  &0  &0  &0  &0    \\
5   &0  &1  &0  &0  &0  &0  &0  &0    \\
6   &0  &1  &0  &1  &0  &0  &0  &1    \\
7   &0  &1  &1  &0  &0  &0  &1  &0    \\
8   &0  &1  &1  &1  &0  &0  &1  &1    \\
9   &1  &0  &0  &0  &0  &0  &0  &0    \\
10   &1  &0  &0  &1  &0  &1  &0  &0    \\
11   &1  &0  &1  &0  &1  &0  &0  &0    \\
12   &1  &0  &1  &1  &1  &1  &0  &0    \\
13   &1  &1  &0  &0  &0  &0  &0  &0    \\
14   &1  &1  &0  &1  &0  &1  &0  &1    \\
15   &1  &1  &1  &0  &1  &0  &1  &0    \\
16   &1  &1  &1  &1  &1  &1  &1  &1    \\
\hline\hline
\end{tabular}
}
}
\caption{Construction of the correlation polytope for four  events. The 16 possible cases are represented by
the truth table, whose rows can be interpreted as eight-dimensional vectors forming the vertices of the correlation polytope.
 \label{2006-ql-t2}
}
\end{table}

By solving the hull problem, one obtains a set of conditions of possible experience
which represent the bounds on classical probabilities enumerated in Table
 \ref{2006-ql-t3}.
For historical reasons, the bounds 17-18, 19-20, 21-22, and 23-24
are  called the Clauser-Horne inequalities \cite{cl-horne,clauser}.
They are equivalent (up to permutations of $p_i,q_i$),
and are the only additional inequalities structurally different from the two-event ``1--1''
case.
\begin{table}
\centerline{
{\footnotesize
\begin{tabular}{|c|cc|}
\hline\hline
 & full facet inequality &  inequality for $p_1=p_2=q_1=q_2={1\over 2}$\\
\hline
 1 &     $p_1q_1 \ge 0$  &
 $ p_1q_1\ge 0$   \\
 2  &     $p_1q_2 \ge 0$  &
 $ p_1q_2\ge 0$   \\
 3  &     $p_2q_1 \ge 0$  &
 $ p_2q_1\ge 0$   \\
 4  &     $p_2q_2 \ge 0$  &
 $ p_2q_2\ge 0$   \\
 5 &     $p_1 \ge   p_1q_1$  &
 ${1\over 2} \ge   p_1q_1$   \\
 6  &     $p_1 \ge   p_1q_2$  &
 ${1\over 2} \ge   p_1q_2$   \\
 7  &     $q_1 \ge   p_1q_1$  &
 ${1\over 2} \ge   p_1q_1$   \\
 8  &     $q_1 \ge   p_1q_2$  &
 ${1\over 2} \ge   p_1q_2$   \\
 9  &     $p_2 \ge   p_2q_1$  &
 ${1\over 2} \ge   p_2q_1$   \\
 10  &     $p_2 \ge   p_2q_2$  &
 ${1\over 2} \ge   p_2q_2$   \\
 11  &     $q_2 \ge   p_2q_1$  &
 ${1\over 2} \ge   p_2q_1$   \\
 12  &     $q_2 \ge   p_1q_2$  &
 ${1\over 2} \ge   p_2q_2$   \\
 13  &     $p_1q_1 \ge   p_1 + q_1 - 1$  &
 $p_1q_1 \ge  0$   \\
 14 &     $p_1q_2 \ge   p_1 + q_2 - 1$  &
 $p_1q_2 \ge  0$   \\
 15 &     $p_2q_1 \ge   p_2 + q_1 - 1$  &
 $p_2q_1 \ge   0$   \\
 16 &     $p_2q_2 \ge   p_2 + q_2 - 1$  &
 $p_2q_2 \ge  0 $   \\
 17 &     $0 \ge   p_1q_1 + p_1q_2 + p_2q_1 - p_2q_2 - p_1 - q_1$  &
 $1 \ge  + p_1q_1 + p_1q_2 + p_2q_1 - p_2q_2$   \\
 18 &     $p_1q_1 + p_1q_2 + p_2q_1 - p_2q_2 - p_1 - q_1 \ge -1$  &
 $ p_1q_1 + p_1q_2 + p_2q_1 - p_2q_2 \ge 0$   \\
 19 &     $0 \ge   p_1q_1 + p_1q_2 - p_2q_1 + p_2q_2- p_1 - q_2$  &
 $1 \ge  + p_1q_1 + p_1q_2 - p_2q_1 + p_2q_2$   \\
 20  &     $p_1q_1 + p_1q_2 - p_2q_1 + p_2q_2- p_1 - q_2 \ge -1$  &
 $p_1q_1 + p_1q_2 - p_2q_1 + p_2q_2 \ge 0$   \\
 21 &     $ 0\ge p_1q_1 - p_1q_2 + p_2q_1 + p_2q_2- p_2 - q_1$  &
 $1 \ge   p_1q_1 - p_1q_2 + p_2q_1 + p_2q_2$   \\
 22&     $p_1q_1 - p_1q_2 + p_2q_1 + p_2q_2- p_2 - q_1 \ge -1$  &
 $p_1q_1 - p_1q_2 + p_2q_1 + p_2q_2 \ge 0$   \\
 23  &     $0 \ge   - p_1q_1 + p_1q_2 + p_2q_1 + p_2q_2- p_2 - q_2$  &
 $1 \ge  - p_1q_1 + p_1q_2 + p_2q_1 + p_2q_2$   \\
 24  &     $- p_1q_1 + p_1q_2 + p_2q_1 + p_2q_2- p_2 - q_2 \ge -1$  &
 $  - p_1q_1 + p_1q_2 + p_2q_1 - p_2q_2 \ge 0$   \\
 \hline\hline
\end{tabular}
}
}
\caption{Construction of the correlation polytope for four  events.
The 24 faces of the polytope spanned by the vertices corresponding to the rows
enumerated in Table~\protect\ref{2006-ql-t2}.
The bounds 17-18, 19-20, 21-22, and 23-24 are the Clauser-Horne inequalities.
 \label{2006-ql-t3}
}
\end{table}

\subsubsection{Six event ``3--3'' case}
A similar calculation \cite{2000-poly} for six events $E_1,E_2,E_3,F_1,F_2,F_3$
depicted in Fig.~\ref{2006-ql-2} yields an additional
independent \cite{collins-gisin-2003,sliwa-2003} inequality
for their probabilities
$p_1,p_2,p_3,q_1,q_2,q_3$ and their joint probabilities of the type
$$
p_1q_1+p_2q_2+p_1q_3+p_2q_1+p_2q_2-p_2q_3+p_3q_1-p_3q_2
\le
p_1 +2q_1+q_2.
$$


\section{Quantum contexts}

Omniscience in a classical sense is no longer possible for quantum systems.
Some of the reasons are:
(i) quantum complementarity  and,
algebraically associated with it, the breakdown of distributivity;
(ii)  the impossibility to consistently assign truth and falsity for all
observables simultaneously and, associated with it,
the nonexistence of two-valued measures on even finite subsets of Hilbert logics;
and
(iii) the alleged randomness of certain single outcomes.

\subsection{Hilbert lattices as quantum logics}
\index{Hilbert lattices}

Quantum logic has been introduced by
Garrett Birkhoff and John von Neumann \cite{v-neumann-49,birkhoff-36,ma-57,jauch,pulmannova-91} in the thirties.
They  organized it {\em top-down},
starting from the Hilbert space
formalism of quantum mechanics.
Certain entities of Hilbert spaces are identified
with propositions,
partial order relations and lattice operations.
These relations and operations are
identified with the logical
implication relation and operations such as ``and,'' ``or,'' and the negation.
Thereby, as we shall see, the resulting logical structures are ``nonclassical,'' in particular ''nonboolean.''

Kochen and Specker \cite{kochen2,kochen3} suggested to consider only relations and operations among compatible, co-measurable observables;
i.e., within Boolean subalgebras, which will be identified with blocks and contexts
of Hilbert lattices.
Nevertheless, some of their theorems formally take into account ensembles of contexts \cite{kochen1} for which a multitude of incompatible observables contribute.

If theoretical physics is assumed to be a faithful
representation of our experience, such an ``empirical,'' ``operational''
\cite{bridgman27,bridgman,bridgman52}
logic derives
its justification by the phenomena themselves.
In this sense,
one of the main justifications for quantum logic is the
construction of the logical and algebraic order of events based on empirical findings.

\subsubsection{Definition}

The dimensionality of the Hilbert space for a given quantum system
depends on the number of possible mutually exclusive outcomes.
In the spin--${1\over 2}$ case, for example, there are two outcomes
``up'' and ``down,'' associated with spin state measurements along arbitrary directions.
Thus, the dimensionality of Hilbert space needs to be two.

Then the following identifications can be made.
Table
 \ref{tcompa} lists the identifications of relations of operations of
classical Boolean set-theoretic and quantum Hillbert lattice types.
\begin{table}
\begin{center}
{\footnotesize
 \begin{tabular}{|ccccc|} \hline\hline
 generic lattice  &  order relation   & ``meet''
&
``join''  & ``complement''\\
\hline
propositional&implication&disjunction&conjunction&negation\\
calculus&$\rightarrow$&``and'' $\wedge$&``or'' $\vee$&``not'' $\neg$\\
\hline
``classical'' lattice  &  subset $\subset $  & intersection $\cap$ &
union
$\cup$ & complement\\
of subsets&&&&\\
of a set&&&&\\
\hline
Hilbert & subspace& intersection of & closure of     & orthogonal \\
lattice & relation& subspaces $\cap$&  linear& subspace   \\
        & $\subset$ &                 & span $\oplus$  &  $\perp$   \\
\hline
lattice of& $E_1E_2=E_1$& $E_1E_2$& $E_1+E_2-E_1E_2$& orthogonal\\
commuting&&&&projection\\
\{noncommuting\}&&\{$\displaystyle\lim_{n\rightarrow \infty}(E_1E_2)^n$\}&&\\
projection&&&&\\
operators&&&&\\
 \hline\hline
 \end{tabular}
}
 \caption{Comparison of the identifications of lattice relations and
 operations for the lattices of subsets of a set, for
 experimental propositional calculi, for  Hilbert lattices, and for
lattices of commuting projection operators.
 \label{tcompa}}
 \end{center}
\end{table}

\begin{description}
\item[$\bullet$]
Any closed linear
subspace of --- or, equivalently, any
projection operator on ---  a Hilbert space corresponds to an elementary
proposition. The elementary {``true''}--{``false''} proposition can in
English be spelled out explicitly as
\begin{quote}
``The physical system has a property corresponding to the associated
closed linear subspace.''
\end{quote}

\item[$\bullet$]
The logical {``and''} operation is identified with the set
theoretical intersection of two propositions ``$\cap$''; i.e., with the
intersection of two subspaces.
It is denoted by the symbol ``$\wedge$''.
So, for two
propositions $p$ and $q$ and their associated closed linear
subspaces
${\mathfrak M}_p$ and
${\mathfrak M}_q$,
$$
{\mathfrak M}_{p\wedge q} = \{x \mid x\in
{\mathfrak M}_p, \;
x\in {\mathfrak M}_q\} .$$

\item[$\bullet$]
The logical {``or''} operation is identified with the closure of the
linear span ``$\oplus$'' of the subspaces corresponding to the two
propositions.
 It is denoted by the symbol ``$\vee$''.
So, for two
propositions $p$ and $q$ and their associated closed linear
subspaces
${\mathfrak M}_p$ and
${\mathfrak M}_q$,
$$
{\mathfrak M}_{p\vee q} =
{\mathfrak M}_{p} \oplus
{\mathfrak M}_{q} =
 \{x \mid x=\alpha y+\beta z,\; \alpha,\beta \in {\mathbb C},\; y\in
{\mathfrak M}_p, \;
z\in {\mathfrak M}_q\} .$$

The symbol $\oplus$ will used to indicate the closed linear subspace
spanned by two vectors. That is,
$$u\oplus v=\{ w\mid w=\alpha u+ \beta v,\; \alpha,\beta \in {\mathbb C}
,\; u,v \in {\mathfrak H}\}.$$

Notice that
a vector of Hilbert space may be an element of
$
{\mathfrak M}_{p} \oplus
{\mathfrak M}_{q}
$
without being an element of either
$
{\mathfrak M}_{p} $ or
${\mathfrak M}_{q}
$, since
$
{\mathfrak M}_{p} \oplus
{\mathfrak M}_{q}
$
includes all the vectors in
$
{\mathfrak M}_{p} \cup
{\mathfrak M}_{q}
$, as well as all of their linear combinations (superpositions) and
their limit vectors.

\item[$\bullet$]
The logical {``not''}-operation, or ``negation'' or ``complement,''
is
identified with operation of taking the orthogonal subspace ``$\perp$''.
It is denoted by the symbol ``~$'$~''.
In particular, for a
proposition $p$ and its associated closed linear
subspace
${\mathfrak M}_p$, the negation $p'$ is associated with
$$
{\mathfrak M}_{p'} =
 \{x \mid (x,y)=0,\; y\in
{\mathfrak M}_p
\} ,$$
where $(x,y)$ denotes the scalar product of $x$ and $y$.

\item[$\bullet$]
The logical {``implication''} relation is identified with the set
theoretical subset relation ``$\subset$''.
It is denoted by the symbol ``$\rightarrow$''.
So, for two
propositions $p$ and $q$ and their associated closed linear
subspaces
${\mathfrak M}_p$ and
${\mathfrak M}_q$,
$$
{p\rightarrow q} \Longleftrightarrow
{\mathfrak M}_{p} \subset
{\mathfrak M}_{q}.$$

\item[$\bullet$]
A trivial statement which is always {``true''} is denoted by $1$.
It is represented by the entire Hilbert space $\mathfrak H$.
So, $${\mathfrak M}_1=\mathfrak H.$$

\item[$\bullet$]
An absurd statement which is always {``false''} is denoted by $0$.
It is represented by the zero vector $0$.
So, $${\mathfrak M}_0= 0.$$
\end{description}

\subsubsection{Diagrammatical representation, blocks, complementarity}

Propositional structures are often represented by
Hasse and Greechie diagrams.
\index{Hasse diagram}
\index{Greechie diagram}
A {\em Hasse diagram} is a convenient representation of the
logical implication,
as well as of the {``and''} and {``or''}
operations
among propositions.
 Points
``~$\bullet$~'' represent propositions. Propositions
which are implied by other ones are drawn higher than the other ones.
Two propositions are connected by a line if one implies the other.
Atoms are propositions which ``cover'' the least element $0$; i.e.,
they lie ``just above'' $0$ in a Hasse diagram of the partial order.

A much more compact representation of the propositional calculus can be
given in terms of
its {\em Greechie diagram} \cite{greechie:71}.
In this representation, the emphasis is on Boolean subalgebras.
Points ``~$\circ$~'' represent the atoms.
\index{Greechie diagram}
If they belong to the same Boolean subalgebra, they are connected by edges or smooth curves.
The collection of all atoms and elements belonging to the same Boolean subalgebra is called {\em block};
\index{block}
i.e., every block represents a Boolean subalgebra within a nonboolean structure.
The blocks can be joined or pasted together as follows.
\begin{description}
\item[$\bullet$]
The tautologies of all blocks are identified.
\item[$\bullet$]
The absurdities of all blocks are identified.
\item[$\bullet$]
Identical elements in different blocks are identified.
\item[$\bullet$]
The logical and algebraic structures of all blocks remain intact.
\end{description}
This construction is often referred to as {\em pasting} construction.
If the blocks are only pasted together at the tautology and
the absurdity, one calls the resulting logic a {\em horizontal
sum}.

Every single block represents some ``maximal collection of co-measurable observables''
which will be identified with some quantum {\em context}.
Hilbert lattices can be thought of as the pasting of a continuity of such blocks or contexts.

Note that whereas all propositions within a given block or context are co-measurable;
propositions belonging to different blocks are not.
This latter feature is an expression of  complementarity.
Thus from a strictly operational point of view,
it makes no sense to speak of the ``real physical existence'' of different contexts,
as knowledge of a single context makes impossible the measurement of all the other ones.

Einstein-Podolski-Rosen (EPR) type arguments \cite{epr} utilizing a configuration
sketched in Fig.~\ref{2006-ql-2}
claim to be able to infer two different contexts counterfactually.
One context is measured on one side of the setup, the other context on the other side of it.
By the uniqueness property \cite{svozil-2004-vax,svozil-2006-uniquenessprinciple} of certain two-particle states,
knowledge of a property of one particle entails the certainty
that, if this property were measured on the other particle as well, the outcome of the measurement would be
a unique function of the outcome of the measurement performed.
This makes possible the measurement of one context, as well as the simultaneous counterfactual inference of another, mutual exclusive, context.
Because, one could argue, although one has actually measured on one side a different, incompatible context compared to the context measured on the other side,
if on both sides the same  context {\em would be measured}, the outcomes on both sides {\em would be uniquely correlated}.
Hence measurement of one context per side is sufficient, for the outcome could be counterfactually inferred on the other side.

As problematic as counterfactual physical reasoning may appear from an operational
point of view even for a two particle state, the simultaneous ``counterfactual inference'' of three or more blocks or contexts fails
because of the missing uniqueness property \cite{svozil-2006-uniquenessprinciple}
of quantum states.

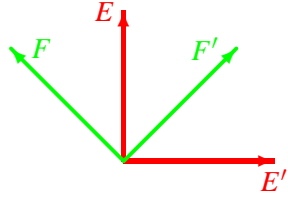
\begin{figure}
\begin{center}
\unitlength .5mm 
\allinethickness{1.5pt} 
\ifx\plotpoint\undefined\newsavebox{\plotpoint}\fi 
\begin{picture}(69.971,40.225)(0,0)
\put(29.971,.225){\color{red}\vector(0,1){40}}
\put(29.971,.225){\color{red}\vector(1,0){40}}
\put(29.971,.225){\color{green}\vector(1,1){30}}
\put(29.943,.225){\color{green}\vector(-1,1){30}}
\put(24.971,39.895){\color{red}\makebox(0,0)[cc]{$E$}}
\put(69.971,-4.775){\color{red}\makebox(0,0)[cc]{$E'$}}
\put(51.641,30.225){\color{green}\makebox(0,0)[cc]{$F'$}}
\put(8.273,30.225){\color{green}\makebox(0,0)[]{$F$}}
\end{picture}
\end{center}
\caption{Two-dimensional configuration of spin $1/2$ state measurements along two directions  ${\bf a}$ and ${\bf b}$.
 \label{2006-ql-nondist}
}
\end{figure}

As a first example, we shall paste together observables of the spin
one-half systems.
We have associated a propositional system
$$L({\bf a})= \{ 0, E, E', 1 \}, $$
corresponding to the outcomes of a measurement of the spin states
along some arbitrary direction ${\bf a}$.
If the spin states would be measured along a different spatial
direction, say
${\bf b}\neq \pm {\bf a} $, an identical propositional
system
$$L({\bf b})= \{ 0, F, F', 1 \} $$
would have resulted, with the propositions $E$ and $F$ explicitly expressed before.
The  two-dimensional Hilbert space representation of this configuration is depicted in
Figure~\ref{2006-ql-nondist}.

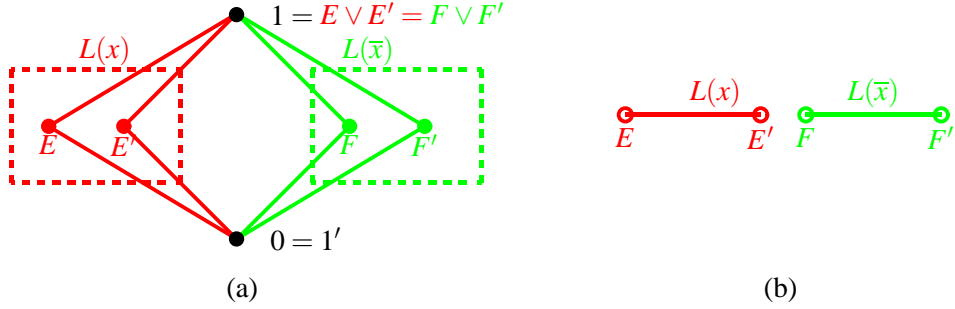
\begin{figure}
\begin{center}
\begin{tabular}{ccc}
\unitlength 0.50mm
\allinethickness{1.5pt} 
\begin{picture}(125.00,60.73)
\put(10.00,30.00){\color{red}\circle{1.5}}
\put(30.00,30.00){\color{red}\circle{1.5}}
\put(90.00,30.00){\color{green}\circle{1.5}}
\put(110.00,30.00){\color{green}\circle{1.5}}
\put(10.00,30.00){\color{red}\circle{3}}
\put(30.00,30.00){\color{red}\circle{3}}
\put(90.00,30.00){\color{green}\circle{3}}
\put(110.00,30.00){\color{green}\circle{3}}
\put(60.00,0.00){\color{red}\line(-1,1){30.00}}
\put(30.00,30.00){\color{red}\line(1,1){30.00}}
\put(60.00,60.00){\color{green}\line(1,-1){30.00}}
\put(90.00,30.00){\color{green}\line(-1,-1){30.00}}
\put(69.00,0.00){\makebox(0,0)[lc]{$0=1'$}}
\put(69.00,60.00){\makebox(0,0)[lc]{$1=\color{red}E\vee E'=\color{green} F \vee F'$}}
\put(30.00,25.00){\makebox(0,0)[cc]{\color{red}$E'$}}
\put(90.00,25.00){\makebox(0,0)[cc]{\color{green}$F$}}
\put(60.00,0.00){\color{red}\line(-5,3){50.00}}
\put(10.00,30.00){\color{red}\line(5,3){50.00}}
\put(60.00,60.00){\color{green}\line(5,-3){50.00}}
\put(110.00,30.00){\color{green}\line(-5,-3){50.00}}
\put(10.00,25.00){\makebox(0,0)[cc]{\color{red}$E$}}
\put(110.00,25.00){\makebox(0,0)[cc]{\color{green}$ F'$}}
\put(25.00,50.00){\makebox(0,0)[cc]{\color{red}$L(x)$}}
\put(95.00,50.00){\makebox(0,0)[cc]{\color{green}$L(\overline x)$}}
\put(0.00,15.00){\color{red}\dashbox{2.00}(45.00,30.00)[cc]{}}
\put(80.00,15.00){\color{green}\dashbox{2.00}(45.00,30.00)[cc]{}}
\put(60.00,0.00){\circle*{3}}
\put(60.00,59.67){\circle*{3}}
\end{picture}
&
\qquad
\qquad
\qquad
&
\unitlength 0.30mm
\allinethickness{1.5pt} 
\begin{picture}(141.06,65.00)
\put(0.00,55.00){\color{red}\circle{6}}
\put(60.00,55.00){\color{red}\circle{6}}
\put(0.00,45.00){\color{red}\makebox(0,0)[cc]{$E$}}
\put(60.00,45.00){\color{red}\makebox(0,0)[cc]{$E'$}}
\put(0.00,55.00){\color{red}\line(1,0){60.00}}
\put(80.00,55.00){\color{green}\circle{6}}
\put(140.00,55.00){\color{green}\circle{6}}
\put(80.00,45.00){\color{green}\makebox(0,0)[cc]{$ F $}}
\put(140.00,45.00){\color{green}\makebox(0,0)[cc]{$ F'$}}
\put(80.00,55.00){\color{green}\line(1,0){60.00}}
\put(40.00,65.00){\color{red}\makebox(0,0)[cc]{$L(x)$}}
\put(110.00,65.00){\color{green}\makebox(0,0)[cc]{$L(\overline x)$}}
\end{picture}
\\
(a)&&(b)
\end{tabular}
\end{center}
\caption{\label{f-hd-mo2}
(a)
Hasse diagram of the ``Chinese lantern'' form obtained by the pasting of
two spin one-half propositional systems
$L(x)$ and
$L(\overline x)$ which are nonco-measurable.
The resulting logical structure is a modular orthocomplemented lattice
$L(x)\oplus L(\overline{x}) = MO_2$.
\index{$MO_2$}
The blocks (without $0,1$) are indicated by dashed boxes.
(b) Greechie
diagram  of the configuration depicted in (a).
}
\end{figure}

$L({\bf a})$ and $L({\bf b})$ can be joined by pasting them together.
In particular, we identify their tautologies and absurdities; i.e., $0$ and $1$.
All the other propositions remain distinct.
We then obtain a propositional structure
$$L({\bf a})\oplus L({\bf b}) =
MO_2$$
whose Hasse diagram is of the ``Chinese lantern'' form and is drawn
\index{chinese lantern}
in Figure \ref{f-hd-mo2}(a). The corresponding Greechie Diagram is drawn in
Figure \ref{f-hd-mo2}(b).
Here, the ``$O$'' stands for {\em orthocomplementation,}
expressing the fact that for every element there exists an orthogonal complement.
The term ``$M$''
stands for {\em modularity}; i.e., for all $x\rightarrow b$, $x \vee (a\wedge b)=(x\vee a)\wedge b$.
The subscript ``2'' stands for the pasting of
two Boolean subalgebras $2^2$.
Since all possible directions ${\bf a}\in {\mathbb R}^3$ form a continuum, the Hilbert lattice is a continuum of pastings of subalgebras of the form $L({\bf a})$.

The propositional system obtained is  not a
classical Boolean algebra,
since the distributive laws are not satisfied; i.e.,
\begin{eqnarray*}
\begin{array}[b]{ccc}
F \vee (E \wedge E')&\stackrel{?}{=}&(F\vee E) \wedge (F \vee E')\\
F \vee 0&\stackrel{?}{=}&1 \wedge 1\\
F &\neq &1,
\end{array}
\\
\begin{array}[b]{ccc}
F \wedge (E \vee E')&\stackrel{?}{=}&(F\wedge E) \vee (F \wedge E')\\
F \wedge 1&\stackrel{?}{=}&0 \vee 0\\
F &\neq &0.
\end{array}
\end{eqnarray*}

Notice that the expressions can be easily evaluated by using the Hasse
diagram
\ref{f-hd-mo2}(a):
For any $a,b$,
$a\vee b$ is just the least element which is connected
by $a$ and $b$;
$a\wedge b$ is just the highest element connected
to $a$ and $b$. Intermediates which are not connected to both $a$
and $b$ do not count. That is,
\begin{center}
\unitlength .5mm 
\allinethickness{1pt} 
\ifx\plotpoint\undefined\newsavebox{\plotpoint}\fi 
\begin{picture}(155.095,20.945)(0,0)
\put(0,0){\line(1,1){20}}
\put(20,20){\line(1,-1){20}}
\put(110.095,20){\line(1,-1){20}}
\put(130.095,0){\line(1,1){20}}
\put(150.095,20){\circle*{1.89}}
\put(110.095,20){\circle*{1.89}}
\put(20,20){\circle*{1.89}}
\put(130.095,0){\circle*{1.89}}
\put(40,0){\circle*{1.89}}
\put(0,0){\circle*{1.89}}
\put(5,0){\makebox(0,0)[cc]{$a$}}
\put(45,0){\makebox(0,0)[cc]{$b$}}
\put(115.095,20){\makebox(0,0)[cc]{$a$}}
\put(155.095,20){\makebox(0,0)[cc]{$b$}}
\put(25,20){\makebox(0,0)[lc]{$a\vee b$}}
\put(135.425,0){\makebox(0,0)[lc]{$a \wedge b$}}
\end{picture}
\end{center}
$a\vee b$ is called a least upper bound of $a$ and $b$.
\index{least upper bound}
$a\wedge b$ is called a greatest lower bound of $a$ and $b$.
\index{greatest lower bound}

$MO_2$ is a specific example of an algebraic structure which is called a
{\em lattice}.
\index{lattice}
Any two elements of a lattice have a least upper and a greatest lower
bound satisfying the commutative, associative and absorption laws.

Nondistributivity is the algebraic expression of nonclassicality,
but what is the algebraic reason for nondistributivity?
It is, heuristically speaking, scarcity, the lack of necessary algebraic elements to ``fill up''
all propositions necessary to obtain one and the same result
in both ways as expressed by the distributive law.

\subsection{Quantum contexts as blocks}

All that is operationally knowable for a given quantized system is a {\em single block}
representing co-measurable observables.
Thus, single blocks or, in another terminology, maximal Boolean subalgebras of Hilbert lattices,
will be identified with quantum contexts.
As Hilbert lattices are pastings of a continuity of blocks or contexts, contexts are the building blocks of quantum logics.

A quantum context can equivalently be formalized
by a single (nondegenerate) ``maximal'' self-adjoint operator  $C$, such that all commuting, compatible co-measurable observables are functions thereof.
(e.g., Ref.~\cite{v-neumann-49}, Sec.~II.10, p. 90, English translation p.~173; Ref.~\cite{kochen1}, \S~2; Ref.~\cite{neumark-54}, pp.~227,228;  Ref.~\cite{halmos-vs}, \S~84).
Note that mutually commuting opators have identical pairwise orthogonal sets of eigenvectors (forming an orthonormal basis)
which correspond to pairwise orthogonal projectors adding up to unity.
The spectral decompositions of the mutually commuting opators thus contain sums of identical pairwise orthogonal projectors.

Thus the ``maximal'' self-adjoint operator  ${C}$ has a spectral
decomposition into some complete set of orthogonal projectors ${E}_i$
which correspond to elementary ``yes''-``no'' propositions in the Von Neumann-Birkhoff type sense \cite{v-neumann-49,birkhoff-36}.
That is, ${C}=\sum_{i=1}^n c_i { E}_i$
with mutually different $c_i$ and $\sum_{i=1}^n { E}_i= {\mathbb I}$.
In $n$ dimensions, contexts can be viewed as $n$-pods spanned by the $n$ orthogonal vectors corresponding to the projectors
${ E}_1, { E}_2, \cdots,{ E}_n$.
As there exist many such representations with many different sets of coefficients $c_i$, ``maximal'' operator are not unique.

An observable belonging to two or more contexts is called {\em link observable}.
\index{link observable}
Contexts can thus be depicted by Greechie  diagrams
\cite{greechie:71},  consisting of {\em points} which
symbolize observables (representable by the spans of vectors
in $n$-dimensional Hilbert space).
Any $n$ points belonging to a context; i.e., to a maximal set of co-measurable observables
(representable as some orthonormal basis of  $n$-dimensional Hilbert space),
are connected by {\em smooth curves}.
Two smooth curves may be crossing in  common {\em link observables}.
In three dimensions, smooth curves and the associated points stand for tripods.
Still another compact representation is in terms of Tkadlec diagrams \cite{tkadlec-00},
in which points represent complete tripods and smooth curves represent
single legs interconnecting them.

In two dimensional Hilbert space, interlinked contexts do not exist,
since every context is fixed by the assumption of one property.
The entire context is just this property, together with its negation,
which corresponds to the orthogonal ray (which spans a one dimensional subspace)
or projection associated with the ray corresponding to the property.

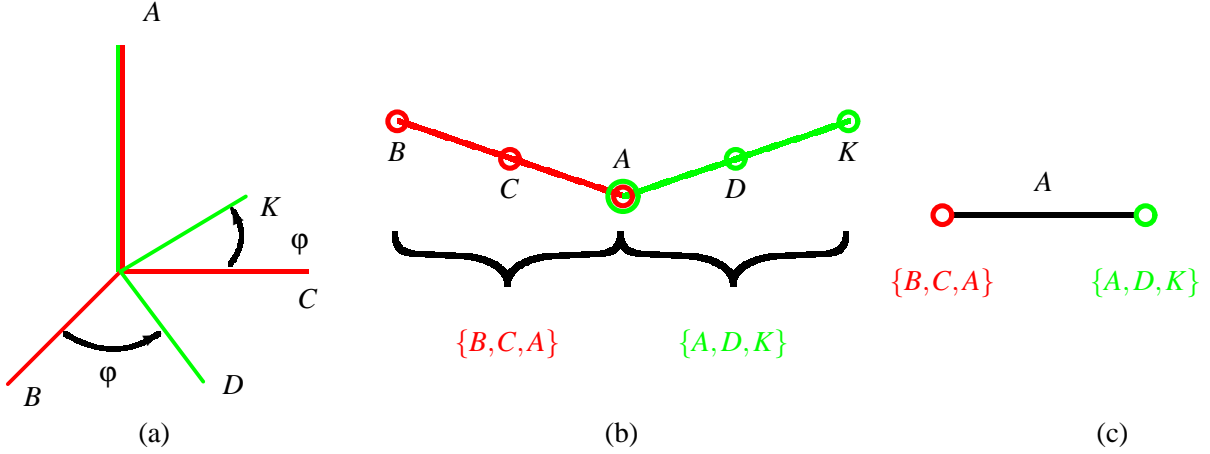
\begin{figure}
\begin{tabular}{ccccc}
\unitlength 1mm
\allinethickness{1.5pt} 
\begin{picture}(40.00,49.67)
\put(14.70,45.00){\color{green}\line(0,-1){30.00}}
\put(15.30,45.00){\color{red}\line(0,-1){30.00}}
\put(15.00,15.00){\color{red}\line(-1,-1){15.00}}
\put(15.00,15.00){\color{red}\line(1,0){25.00}}
\put(15.00,15.00){\color{green}\line(3,-4){11.00}}
\put(15.00,15.00){\color{green}\line(5,3){16.67}}
\put(3.33,-1.67){\makebox(0,0)[cc]{$B$}}
\put(30.00,0.00){\makebox(0,0)[cc]{$D$}}
\put(40.00,11.33){\makebox(0,0)[cc]{$C$}}
\put(35.00,23.67){\makebox(0,0)[cc]{$K$}}
\put(19.33,49.67){\makebox(0,0)[cc]{$A$}}
\put(20.00,6.67){\vector(2,1){0.2}}
\bezier{60}(7.67,6.67)(14.33,2.67)(20.00,6.67)
\put(30.00,23.00){\vector(-1,2){0.2}}
\bezier{36}(29.67,16.00)(32.33,19.67)(30.00,23.00)
\put(13.33,1.00){\makebox(0,0)[cc]{$\varphi$}}
\put(40.00,18.67){\makebox(0,0)[rc]{$\varphi$}}
\end{picture}
&
\qquad
\qquad
&
\unitlength 1mm
\allinethickness{2pt} 
\begin{picture}(61.33,36.00)
\multiput(0.33,35.00)(0.36,-0.12){84}{\color{red}\line(1,0){0.36}}
\multiput(30.33,25.00)(0.36,0.12){84}{\color{green}\line(1,0){0.36}}
\put(0.33,35.00){\color{red}\circle{2.50}}
\put(15.33,30.00){\color{red}\circle{2.50}}
\put(30.33,25.00){\color{red}\circle{2.50}}
\put(30.33,25.00){\color{green}\circle{4.00}}
\put(45.33,30.00){\color{green}\circle{2.50}}
\put(60.33,35.00){\color{green}\circle{2.50}}
\put(60.33,31.00){\makebox(0,0)[cc]{$K$}}
\put(45.33,26.00){\makebox(0,0)[cc]{$D$}}
\put(30.33,30.00){\makebox(0,0)[cc]{$A$}}
\put(15.33,26.00){\makebox(0,0)[cc]{$C$}}
\put(0.33,31.00){\makebox(0,0)[cc]{$B$}}
\bezier{24}(0.00,20.00)(0.00,17.33)(3.00,17.33)
\bezier{28}(3.00,17.33)(10.00,17.00)(10.00,17.00)
\bezier{32}(10.00,17.00)(15.00,16.00)(15.00,13.33)
\bezier{24}(30.00,20.00)(30.00,17.33)(27.00,17.33)
\bezier{28}(27.00,17.33)(20.00,17.00)(20.00,17.00)
\bezier{32}(20.00,17.00)(15.00,16.00)(15.00,13.33)
\put(15.00,5.33){\color{red}\makebox(0,0)[cc]{$\{B,C,A\}$}}
\bezier{24}(60.00,20.00)(60.00,17.33)(57.00,17.33)
\bezier{28}(57.00,17.33)(50.00,17.00)(50.00,17.00)
\bezier{32}(50.00,17.00)(45.00,16.00)(45.00,13.33)
\bezier{24}(30.00,20.00)(30.00,17.33)(33.00,17.33)
\bezier{28}(33.00,17.33)(40.00,17.00)(40.00,17.00)
\bezier{32}(40.00,17.00)(45.00,16.00)(45.00,13.33)
\put(45.00,5.33){\color{green}\makebox(0,0)[cc]{$\{A,D,K\}$}}
\end{picture}
&
\qquad
\qquad
&
\unitlength 0.90mm
\allinethickness{2pt} 
\begin{picture}(51.37,10.00)
\put(1.33,25.00){\line(1,0){27.33}}
\put(0.00,25.00){\color{red}\circle{2.75}}
\put(30.00,25.00){\color{green}\circle{2.75}}
\put(15.00,30.00){\makebox(0,0)[cc]{$A$}}
\put(30.00,15.00){\color{green}\makebox(0,0)[cc]{$\{A,D,K\}$}}
\put(0.00,15.00){\color{red}\makebox(0,0)[cc]{$\{B,C,A\}$}}
\end{picture}
\\
(a)&&(b)&&(c)\\
\end{tabular}
\caption{Three equivalent representations of the same geometric configuration:
(a) Two tripods with a common leg;
(b) Greechie (orthogonality) diagram: points stand for individual basis vectors, and
orthogonal tripods are drawn as smooth curves;
(c) Tkadlec diagram: points represent complete tripods and smooth curves represent
single legs interconnecting them.
\label{2004-qnc-f1}}
\end{figure}

The simplest nontrivial configuration of interlinked contexts exists in three-dimensional Hilbert space.
Consider an arrangement
of five observables $A$, $B$, $C$, $D$, $K$ with two systems
of operators
$\{A,B,C\}$
and
$\{D,K,A\}$,
the  contexts,
which are interconnected by $A$.
Within a context, the operators commute and the associated observables are co-measurable.
For two different contexts, operators outside the link operators do not commute.
$A$ is a link observable.
This propositional structure (also known as $L_{12}$) can be represented in three-dimensional Hilbert space
by two tripods with a single common leg.
Fig.~\ref{2004-qnc-f1} depicts this configuration in three-dimensional real vector space,
as well as in the associated Greechie and Tkadlec diagrams.
The operators  $B,C,A$ and $D,K,A$ can be identified with the projectors corresponding
to the two bases
$$
\begin{array}{lcl}
B_{B-C-A}&=&
\{
(1,0,0)^T,
(0,1,0)^T,
(0,0,1)^T
\}
,
\\
B_{D-K-A}&=&
\{
(\cos \varphi , \sin \varphi ,0)^T,
(-\sin \varphi ,\cos \varphi , 0)^T,
(0,0,1)^T
\},
\end{array}
\label{e-vaxjo1}
$$
(the superscript ``$T$'' indicates transposition).
Their matrix representation is the  dyadic product of every vector with itself.

Physically, the union of contexts $\{B,C,A\}$ and $\{D,K,A\}$ interlinked along $A$ does not have any direct
operational meaning; only a single context can be measured along a single quantum at a time;
the other being irretrievably lost if no reconstruction of the original state is possible.
Thus, in a direct way, testing the value of observable $A$ against different
contexts $\{B,C,A\}$ and $\{D,K,A\}$ is metaphysical.

It is, however, possible to counterfactually retrieve information
about the two different contexts of a single quantum indirectly
by considering a singlet state
$
\vert \Psi_2 \rangle
= ({1/ \sqrt{3}})(
\vert + -\rangle
+
\vert - +\rangle
-
\vert 0 0\rangle
)$
via the ``explosion view'' Einstein-Podolsky-Rosen type of argument depicted in Fig.~\ref{2006-ql-2}.
Since the state is form invariant with respect to variations of the measurement angle
and at the same time satisfies the uniqueness property \cite{svozil-2006-uniquenessprinciple},
one may retrieve the first context
$\{B,C,A\}$ from the first quantum
and the second context $\{D,K,A\}$ from the second quantum.
(This is a standard procedure in Bell type arguments with two spin one-half quanta.)

More tightly interlinked contexts such as
$\{A,B,C\}-\{C,D,E\}-\{E,F,A\}$, whose Greechie diagram is a triangle with the edges $A$, $C$ and $E$,
or
$\{A,B,C\}-\{C,D,E\}-\{E,F,G\}-\{G,H,A\}$, whose Greechie diagram is a quadrangle with the edges $A$, $C$, $E$ and $G$,
cannot be represented in Hilbert space and thus have no realization in quantum logics.
The five contexts
$\{A,B,C\}-\{C,D,E\}-\{E,F,G\}-\{G,H,I\}-\{I,J,A\}$
whose Greechie diagrams is a pentagon with the edges $A$, $C$, $E$, $G$ and $I$ have realizations in ${\mathbb R}^3$ \cite{svozil-tkadlec}.

\subsection{Probability theory}

\subsubsection{Kochen-Specker theorem}
\index{Kochen-Specker theorem}

Quantum logics of Hilbert space dimension greater than two
have not a single two-valued state interpretable as consistent, overall truth assignment
\cite{specker-60}.
This is the gist of the beautiful construction of Kochen and Specker \cite{kochen1}.
For similar theorems, see Refs.~\cite{ZirlSchl-65,Alda,Alda2,kamber64,kamber65}.
As a result of the nonexistence of two-valued states, the classical strategy
to construct probabilities by a convex combination of all two-valued states fails entirely.

One of the most compact and comprehensive versions of the Kochen-Specker
proof by contradiction in three-dimensional Hilbert space ${\mathbb R}^3$ has been given
by Peres
\cite{peres-91}.
(For other discussions, see Refs.
\cite{stairs83,redhead,jammer-92,brown,peres-91,peres,penrose-ks,clifton-93,mermin-93,svozil-tkadlec}.)
Peres' version uses
a 33-element
set of lines without a two-valued state.
The direction vectors of these lines
arise by all permutations of coordinates from
$$
(0,0,1),\;
(0,\pm 1,1),\;
(0,\pm 1,\sqrt{2}),\;
\;\textrm{ and }\; (\pm 1,\pm 1,\sqrt{2}).
\label{p-ksl}
$$
These lines can be generated (by
the ``nor''-operation between nonorthogonal propositions)
by the three lines
\cite{svozil-tkadlec}
$$(1,0,0),\; (1,1,0),\; (\sqrt2,1,1).$$
Note that as three arbitrary but mutually nonorthogonal lines generate
a dense set of lines  \cite{havlicek},
it can be expected that any such triple of lines (not just the one
explicitly mentioned) generates a finite set of lines which does not
allow a two-valued probability measure.

The way it is defined, this set of lines is invariant under interchanges
(permutations) of the $x_1,x_2$ and $x_3$ axes, and under a reversal of
the direction of each of these axes.
This symmetry property allows us to assign the probability measure $1$
to some of the rays without loss of generality.
Assignment of
probability measure $0$ to these rays would be equivalent to renaming
the axes, or reversing one of the axes.

The Greechie diagram
of the Peres configuration
is given in Figure~\ref{sk33} \cite{svozil-tkadlec}.
For simplicity, 24~points which
belong to exactly one edge are  omitted.
 The coordinates should be read as follows:
$\bar{1}\rightarrow -1$ and $2\rightarrow \sqrt{2}$; e.g.,
1$\bar{1}$2
 denotes
$\textrm{Sp} (1,-1,\sqrt{2})$.
 Concentric circles indicate the (non
 orthogonal) generators mentioned above.
\begin{figure}
\unitlength .45\textwidth
\newcommand{\emline}[4]{\put(#1,#2){\special{em:moveto}}\put(#3,#4){\special{em:lineto}}}
\newsavebox{\vertex}\savebox{\vertex}{{\unitlength1mm\begin{picture}(0,0)\put(0,0){\circle{1}}\end{picture}}}
\newsavebox{\subdiagram}\savebox{\subdiagram}{
{\unitlength1mm\begin{picture}(0,0)\put(0,0){\circle{2}}\end{picture}}}
\newcommand{\discbig}{\usebox{\subdiagram}}
\newcommand{\emlin}[4]{\put(#1,#2){\special{em:moveto}}\put(#3,#4){\special{em:lineto}}}
\newcommand{\disc}{\usebox{\vertex}}
\newcommand{\place}[6]{\put(#1,#2){\hspace{#3pt}\raisebox{#4pt}{\makebox(0,0)[#5]{$#6$}}}}
 \newcommand{\point}[2]{\put(#1,#2){\disc}}
\newcommand{\onethird}[3]{\axis#1\side#1#2\side#1#3\cross#2#3}
\newcommand{\axis}[4]{\point#1\point#2\point#3\point#4\emlin#1#3\emlin#3{0}{0}}
\newcommand{\side}[8]{\point#5\point#6\point#7\point#8\emlin#1#5\emlin#5#6\emlin#2#8\emlin#8#7}
\newcommand{\cross}[8]{\emlin#1#7\emlin#7#3\emlin#3#5\emlin#2{0}{0}}
\catcode`\!=\active  \def!{\bar1}
\begin{center}
\begin{picture}(2,2)(-1,-1)
\onethird
{{{ 0.100}{-0.995}}{{ 0.050}{-0.747}}{{ 0.000}{-0.500}}{{ 0.000}{-0.250}}}%
{{{ 0.643}{-0.766}}{{ 0.087}{-0.050}}{{ 0.100}{-0.250}}{{ 0.536}{-0.112}}}%
{{{-0.643}{-0.766}}{{-0.087}{-0.050}}{{-0.100}{-0.250}}{{-0.536}{-0.112}}}%
\onethird
{{{ 0.812}{ 0.584}}{{ 0.622}{ 0.417}}{{ 0.433}{ 0.250}}{{ 0.217}{ 0.125}}}%
{{{ 0.342}{ 0.940}}{{-0.000}{ 0.100}}{{ 0.167}{ 0.212}}{{-0.171}{ 0.520}}}%
{{{ 0.985}{-0.174}}{{ 0.087}{-0.050}}{{ 0.267}{ 0.038}}{{ 0.365}{-0.408}}}%
\onethird
{{{-0.912}{ 0.411}}{{-0.672}{ 0.330}}{{-0.433}{ 0.250}}{{-0.217}{ 0.125}}}%
{{{-0.985}{-0.174}}{{-0.087}{-0.050}}{{-0.267}{ 0.038}}{{-0.365}{-0.408}}}%
{{{-0.342}{ 0.940}}{{-0.000}{ 0.100}}{{-0.167}{ 0.212}}{{ 0.171}{ 0.520}}}%
\put (0,0){\circle{0.2}}
\put ( 0.000, 0.100){\discbig}
\put ( 0.050,-0.747){\discbig}
\put ( 0.217, 0.125){\discbig}
\footnotesize
\place { 0.100}{-0.995}{ 0}{-6}{ t}{2!!}
\place { 0.050}{-0.747}{ 6}{ 0}{l }{211}
\place { 0.000}{-0.500}{ 6}{ 0}{l }{01!}
\place { 0.000}{-0.250}{ 2}{ 3}{lb}{011}
\place { 0.000}{ 0.100}{ 0}{-4}{ t}{100}
\place { 0.100}{-0.250}{ 4}{-4}{lt}{2!1}
\place {-0.100}{-0.250}{-4}{-4}{rt}{21!}
\place { 0.643}{-0.766}{ 6}{ 0}{l }{102}
\place { 0.365}{-0.408}{ 6}{ 0}{l }{20!}
\place { 0.087}{-0.050}{ 4}{ 0}{lb}{010}
\place {-0.643}{-0.766}{-6}{ 0}{r }{120}
\place {-0.365}{-0.408}{-6}{ 0}{r }{2!0}
\place {-0.087}{-0.050}{-4}{ 0}{rb}{001}
\place { 0.812}{ 0.584}{ 6}{ 0}{lb}{!!2}
\place { 0.622}{ 0.417}{ 6}{ 0}{lt}{112}
\place {-0.912}{ 0.411}{-6}{ 0}{rb}{!2!}
\place {-0.672}{ 0.330}{-6}{ 0}{rt}{121}
\place { 0.433}{ 0.250}{ 3}{-3}{lt}{1!0}
\place { 0.217}{ 0.125}{-6}{ 1}{r }{110}
\place {-0.433}{ 0.250}{-3}{-3}{rt}{10!}
\place {-0.217}{ 0.125}{ 6}{ 1}{l }{101}
\place { 0.267}{ 0.038}{ 6}{ 0}{l }{1!2}
\place { 0.167}{ 0.212}{ 0}{ 6}{lb}{!12}
\place {-0.267}{ 0.038}{-6}{ 0}{r }{12!}
\place {-0.167}{ 0.212}{ 0}{ 6}{rb}{!21}
\place { 0.985}{-0.174}{ 4}{-6}{ t}{201}
\place { 0.536}{-0.112}{ 0}{-6}{lt}{!02}
\place {-0.985}{-0.174}{-4}{-6}{ t}{210}
\place {-0.536}{-0.112}{ 0}{-6}{rt}{!20}
\place { 0.342}{ 0.940}{ 0}{ 6}{ b}{021}
\place { 0.171}{ 0.520}{-4}{ 4}{rb}{0!2}
\place {-0.342}{ 0.940}{ 0}{ 6}{ b}{012}
\place {-0.171}{ 0.520}{ 4}{ 4}{lb}{02!}
\end{picture}
\end{center}
 \caption{Greechie diagram of a finite subset of the continuum of blocks or contexts embeddable in three-dimensional real Hilbert space
 without a two-valued probability measure
  \protect\cite[Figure 9]{svozil-tkadlec}.
\label{sk33}}
\end{figure}

Let us prove that there is no two-valued probability measure
\cite{svozil-tkadlec,tkadlec-96}.
Due to the symmetry of the problem, we can  choose a particular
coordinate axis such that, without loss of generality,
$P(100)=1$.
Furthermore, we may assume (case 1) that
$P(21\bar{1}) = 1$.
It immediately follows that $P(001) = P(010) =
P(102) = P(\bar{1}20) = 0$.
A second glance shows that $P(20\bar{1}) = 1$, $P(1\bar{1}2) = P(112) = 0$.

Let us now suppose (case 1a)
 that $P(201) = 1$. Then  we obtain $P(\bar{1}12) = P(\bar{1}\bar{1}2)
= 0$. We are forced to accept $P(110)
= P(1\bar{1}0)  = 1$ ---  a contradiction, since $(110)$ and
$(1\bar{1}0)$ are
orthogonal to each other and lie on one edge.

Hence we have to assume (case 1b) that $P(201) = 0$.
This gives immediately
$P(\bar{1}02)=1$ and
$P(211) =0$.  Since $P(01\bar{1})=0$, we obtain $P(2\bar{1}\bar{1})=1$ and thus
$P(120)=0$.
This requires $P(2\bar{1}0)=1$ and therefore $P(12\bar{1})=P(121)=0$.
Observe that $P(210) = 1$, and thus $P(\bar{1}2\bar{1}) = P(\bar{1}21) = 0$.
In the following step, we notice that $P(10\bar{1}) =
P(101) = 1$ ---  a contradiction,
since $(101)$ and $(10\bar{1})$ are
orthogonal to each other and lie on one edge.

Thus we are forced to assume (case 2) that
$P(2\bar{1}1) = 1$. There is no third alternative, since $P(011)=0$ due to the
orthogonality with $(100)$. Now we can repeat the argument for case 1 in
its mirrored form.

The most compact way of deriving the Kochen-Specker theorem in four dimensions has been given by Cabello \cite{cabello-96,cabello-99}.
It is depicted in Fig.~\ref{2007-miracles-ksc}.
\begin{figure}
\begin{center}
\begin{tabular}{cc}
\unitlength .6mm 
\allinethickness{2pt} 
\ifx\plotpoint\undefined\newsavebox{\plotpoint}\fi 
\begin{picture}(134.09,125.99)(0,0)

\multiput(86.39,101.96)(.119617225,-.208133971){209}{{\color{green}\line(0,-1){.208133971}}}
\multiput(86.39,14.96)(.119617225,.208133971){209}{{\color{red}\line(0,1){.208133971}}}
\multiput(36.47,101.96)(-.119617225,-.208133971){209}{{\color{yellow}\line(0,-1){.208133971}}}
\multiput(36.47,14.96)(-.119617225,.208133971){209}{{\color{magenta}\line(0,1){.208133971}}}
\color{blue}\put(86.39,15.21){\color{blue}\line(-1,0){50}}
\put(86.39,101.71){\color{violet}\line(-1,0){50}}
\put(36.34,15.16){\color{magenta}\circle{6}}
\put(36.34,15.16){\color{blue}\circle{4}}
\put(52.99,15.16){\color{blue}\circle{4}}
\put(52.99,15.16){\color{cyan}\circle{6}}
\put(69.68,15.16){\color{blue}\circle{4}}
\put(69.68,15.16){\color{orange}\circle{6}}
\put(86.28,15.16){\color{blue}\circle{4}}
\put(86.28,15.16){\color{red}\circle{6}}
\put(93.53,27.71){\color{red}\circle{4}}
\put(93.53,27.71){\color{orange}\circle{6}}
\put(102.37,43.44){\color{red}\circle{4}}
\put(102.37,43.44){\color{olive}\circle{6}}
\put(111.21,58.45){\color{red}\circle{4}}
\color{green}\put(111.21,58.45){\circle{6}}
\put(102.37,73.47){\color{green}\circle{4}}
\put(102.37,73.47){\color{olive}\circle{6}}
\put(93.53,89.21){\color{green}\circle{4}}
\put(93.53,89.21){\color{cyan}\circle{6}}
\put(86.28,101.76){\color{green}\circle{4}}
\put(86.28,101.76){\color{violet}\circle{6}}
\put(69.68,101.76){\color{violet}\circle{4}}
\put(69.68,101.76){\color{cyan}\circle{6}}
\put(52.99,101.76){\color{violet}\circle{4}}
\put(52.99,101.76){\color{orange}\circle{6}}
\put(36.34,101.76){\color{violet}\circle{4}}
\put(36.34,101.76){\color{yellow}\circle{6}}
\put(29.24,89.21){\color{yellow}\circle{4}}
\put(29.24,89.21){\color{orange}\circle{6}}
\put(20.4,73.47){\color{yellow}\circle{4}}
\put(20.4,73.47){\color{olive}\circle{6}}
\put(11.56,58.45){\color{yellow}\circle{4}}
\put(11.56,58.45){\color{magenta}\circle{6}}

\put(20.4,43.44){\color{magenta}\circle{4}}
\put(20.4,43.44){\color{olive}\circle{6}}
\put(29.24,27.71){\color{magenta}\circle{4}}
\put(29.24,27.71){\color{cyan}\circle{6}}

\color{cyan}
\qbezier(29.2,27.73)(23.55,-5.86)(52.99,15.24)
\qbezier(29.2,27.88)(36.93,75)(69.63,101.91)
\qbezier(52.69,15.24)(87.47,40.96)(93.72,89.27)
\qbezier(93.72,89.27)(98.4,125.99)(69.49,102.06)
\color{orange}
\qbezier(93.57,27.73)(99.22,-5.86)(69.78,15.24)
\qbezier(93.57,27.88)(85.84,75)(53.13,101.91)
\qbezier(70.08,15.24)(35.3,40.96)(29.05,89.27)
\qbezier(29.05,89.27)(24.37,125.99)(53.28,102.06)
\color{olive}
\qbezier(20.15,73.72)(-11.67,58.52)(20.15,43.31)
\qbezier(20.33,73.72)(61.34,93.16)(102.36,73.72)
\qbezier(102.36,73.72)(134.09,58.52)(102.53,43.31)
\qbezier(102.53,43.31)(60.99,23.43)(20.15,43.49)
{\color{black}
\put(30.41,114.02){\makebox(0,0)[cc]{$M$}}
\put(30.41,2.65){\makebox(0,0)[cc]{$A$}}
\put(52.68,114.38){\makebox(0,0)[cc]{$L$}}
\put(52.68,2.3){\makebox(0,0)[cc]{$B$}}
\put(91.93,114.2){\makebox(0,0)[cc]{$J$}}
\put(91.93,2.48){\makebox(0,0)[cc]{$D$}}
\put(69.65,114.38){\makebox(0,0)[cc]{$K$}}
\put(73.65,2.3){\makebox(0,0)[cc]{$C$}}
\put(103.24,94.22){\makebox(0,0)[cc]{$I$}}
\put(17.45,94.22){\makebox(0,0)[cc]{$ N$}}
\put(106.24,22.45){\makebox(0,0)[cc]{$E$}}
\put(17.45,22.45){\makebox(0,0)[cc]{$ R$}}
\put(115.13,77.96){\makebox(0,0)[cc]{$H$}}
\put(8.55,77.96){\makebox(0,0)[cc]{$ O$}}
\put(115.13,38.72){\makebox(0,0)[cc]{$F$}}
\put(10.55,38.72){\makebox(0,0)[cc]{$ Q$}}
\put(120.92,57.98){\makebox(0,0)[l]{$ G$}}
\put(1.77,57.98){\makebox(0,0)[rc]{$  P$}}
}
\put(61.341,9.192){\color{blue}\makebox(0,0)[cc]{$a$}}
\put(102.883,35.355){\color{red}\makebox(0,0)[cc]{$b$}}
\put(102.53,84.322){\color{green}\makebox(0,0)[cc]{$c$}}
\put(60.457,108.01){\color{violet}\makebox(0,0)[cc]{$d$}}
\put(18.031,84.145){\color{yellow}\makebox(0,0)[cc]{$e$}}
\put(18.561,33.057){\color{magenta}\makebox(0,0)[cc]{$f$}}
\put(61.341,39.774){\color{olive}\makebox(0,0)[cc]{$g$}}
\put(72.124,67.882){\color{orange}\makebox(0,0)[cc]{$h$}}
\put(48.79,67.705){\color{cyan}\makebox(0,0)[cc]{$i$}}
\end{picture}
&
\unitlength .6mm 
\allinethickness{2pt} 
\ifx\plotpoint\undefined\newsavebox{\plotpoint}\fi 
\begin{picture}(119.854,112.606)(0,0)
\multiput(86.567,102.137)(.119617225,-.208133971){209}{\line(0,-1){.208133971}}
\multiput(86.567,15.137)(.119617225,.208133971){209}{\line(0,1){.208133971}}
\multiput(36.647,102.137)(-.119617225,-.208133971){209}{\line(0,-1){.208133971}}
\multiput(36.647,15.137)(-.119617225,.208133971){209}{\line(0,1){.208133971}}
\multiput(86.087,73.357)(.12,9.54667){3}{\line(0,1){9.54667}}
\multiput(86.267,73.537)(.200967742,-.119758065){124}{\line(1,0){.200967742}}
\multiput(86.087,73.357)(-.60016129,-.119677419){124}{\line(-1,0){.60016129}}
\multiput(86.087,73.187)(-.1198317308,-.139375){416}{\line(0,-1){.139375}}
\multiput(36.951,15.376)(.20325,.119333333){120}{\line(1,0){.20325}}
\multiput(61.341,29.696)(.212166667,-.119333333){120}{\line(1,0){.212166667}}
\multiput(60.991,29.696)(-.119607843,.350980392){204}{\line(0,1){.350980392}}
\multiput(61.161,29.526)(.119813084,.337009346){214}{\line(0,1){.337009346}}
\put(11.844,58.69){\line(3,2){22.804}}
\multiput(34.648,73.892)(.1144118,1.643){17}{\line(0,1){1.643}}
\multiput(86.62,15.38)(-.1198804598,.1341057471){435}{\line(0,1){.1341057471}}
\multiput(34.472,73.716)(.605488189,-.119708661){127}{\line(1,0){.605488189}}
\put(86.567,101.887){\line(-1,0){50}}
\put(86.567,15.387){\line(-1,0){50}}
\put(86.457,101.937){\color{red}\circle{3.4}}
\put(86.457,101.937){\color{red}\circle{1.3}}
\put(86.457,15.337){\color{violet}\circle{3.4}}
\put(86.457,15.337){\color{violet}\circle{1.3}}
\put(85.927,73.127){\color{olive}\circle{3.4}}
\put(85.927,73.127){\color{olive}\circle{1.3}}
\put(34.486,73.849){\color{cyan}\circle{3.4}}
\put(34.486,73.849){\color{cyan}\circle{1.3}}
\put(111.387,58.627){\color{green}\circle{3.4}}
\put(111.387,58.627){\color{green}\circle{1.3}}
\put(11.737,58.627){\color{magenta}\circle{3.4}}
\put(11.737,58.627){\color{magenta}\circle{1.3}}
\put(36.517,101.937){\color{blue}\circle{3.4}}
\put(36.517,101.937){\color{blue}\circle{1.3}}
\put(36.517,15.337){\color{yellow}\circle{3.4}}
\put(36.517,15.337){\color{yellow}\circle{1.3}}
\put(60.941,29.586){\color{orange}\circle{3.4}}
\put(60.941,29.586){\color{orange}\circle{1.3}}
\put(35.885,112.606){\color{blue}\makebox(0,0)[cc]{$a$}}
\put(86.09,111.722){\color{red}\makebox(0,0)[cc]{$b$}}
\put(119.854,55.861){\color{green}\makebox(0,0)[cc]{$c$}}
\put(86.266,6.364){\color{violet}\makebox(0,0)[cc]{$d$}}
\put(35.885,5.834){\color{yellow}\makebox(0,0)[cc]{$e$}}
\put(3.359,58.689){\color{magenta}\makebox(0,0)[cc]{$f$}}
\put(60.634,22.45){\color{orange}\makebox(0,0)[cc]{$h$}}
\put(93.161,77.074){\color{olive}\makebox(0,0)[cc]{$g$}}
\put(28.814,76.544){\color{cyan}\makebox(0,0)[cc]{$i$}}
\end{picture}
\\
(a)&(b)\\
\end{tabular}
\end{center}
\caption{Greechie diagram of a finite subset of the continuum of blocks or contexts embeddable in
four-dimensional real Hilbert space without a two-valued probability measure  \protect\cite{cabello-96,cabello-99}.
The proof of the Kochen-Specker theorem  uses  nine tightly interconnected contexts
$\color{blue}a=\{A,B,C,D\}$,
$\color{red}b=\{D,E,F,G\}$,
$\color{green}c=\{G,H,I,J\}$,
$\color{violet}d=\{J,K,L,M\}$,
$\color{yellow}e=\{M,N,O,P\}$,
$\color{magenta}f=\{P,Q,R,A\}$,
$\color{orange}g=\{B,I,K,R\}$,
$\color{olive}h=\{C,E,L,N\}$,
$\color{cyan}i=\{F,H,O,Q\}$
consisting of the 18 projectors associated with the one dimensional subspaces spanned by
$ A=(0,0,1,-1)    $,
$ B=(1,-1,0,0)    $,
$ C=(1,1,-1,-1)   $,
$ D=(1,1,1,1)     $,
$  E=(1,-1,1,-1)  $,
$  F=(1,0,-1,0)   $,
$  G=(0,1,0,-1)   $,
$  H=(1,0,1,0)    $,
$  I=(1,1,-1,1)   $,
$ J=(-1,1,1,1)    $,
$ K=(1,1,1,-1)    $,
$ L=(1,0,0,1)     $,
$ M=(0,1,-1,0)    $,
$  N=(0,1,1,0)    $,
$  O=(0,0,0,1)    $,
$  P=(1,0,0,0)    $,
$  Q=(0,1,0,0)    $,
$  R=(0,0,1,1)    $.
(a) Greechie diagram representing atoms by points, and  contexts by maximal smooth, unbroken curves.
(b) Dual Tkadlec diagram representing contexts by filled points, and interconnected contexts are connected by lines.
(Duality means that points represent blocks and maximal smooth curves represent atoms.)
Every observable proposition occurs in exactly two contexts.
Thus, in an enumeration of the four observable propositions of each of the nine contexts,
there appears to be an {\em even} number of true propositions.
Yet, as there is an odd number of contexts,
there should be an {\em odd} number (actually nine) of true propositions.   \label{2007-miracles-ksc} }
\end{figure}

\subsubsection{Gleason's derivation of the Born rule}
\index{Gleason theorem}

In view of the nonexistence of classical two-valued states on even finite superstructures of blocks or contexts associated with quantized systems,
one could still resort to classicality {\em within} blocks or contexts.
According to Gleason's theorem, this is exactly the route, the {\em ``via regia,''} to the quantum probabilities, in particular to the Born rule.

According to the {\em Born rule}, the expectation value
\index{Born rule}
$\langle A\rangle$  of an observable $A$ is the trace of $\rho A$; i.e.,
$\langle A\rangle ={\rm tr}(\rho  A)$.
In particular, if $A$ is a projector $E$ corresponding to an elementary yes-no proposition
{\it ``the system has property Q,''} then $\langle E\rangle ={\rm tr}(\rho  E)$ corresponds
to the probability of that property $Q$ if the system is in state $\rho$.
The equations $\rho ^{2} = \rho$ and ${\rm tr}(\rho  ^{2})=1$
are only valid for pure states, because $\rho $ is not an projector and thus idempotent for mixed states.

It is still possible to ascribe a certain degree of classical probabilistic behaviour to a quantum logic by considering
its block superstructure.
Due to their Boolean algebra, blocks are ``classical mini-universes.''
It is one of the mindboggling features of quantum logic that it can be decomposed into a pasting of blocks.
Conversely, by a proper arrangement of ``classical mini-universes,''
quantum Hilbert logics can be obtained.
This theme is used in quantum probability theory,
in particular by the Gleason and the Kochen-Specker theorems.
In this sense, Gleason's theorem can be understood as the functional analytic generalization of
the generation of all classical probability distributions by a convex sum of the extreme cases.

Gleason's theorem
\cite{Gleason,r:dvur-93,c-k-m,peres,hru-pit-2003,rich-bridge}
is  a derivation of the Born rule
from fundamental assumptions about quantum probabilities, guided by
the quasi--classical; i.e., Boolean, sub-parts of quantum theory.
Essentially, the main assumption required for Gleason's theorem is that {\em within} blocks or contexts, the quantum probabilities behave as classical probabilities;
in particular the sum of probabilities over a complete set of mutually exclusive events add up to unity.
With these quasi--classical provisos, Gleason proved that there is no alternative to
the Born rule for Hilbert spaces of dimension greater than two.

\subsection{Quantum violations of classical probability bounds}

Due to the different form of quantum correlations,
which formally is a consequence of the different way of defining
quantum probabilities, the constraints on classical probabilities
are violated by quantum probabilities.
Quantitatively, this can be investigated \cite{filipp-svo-04-qpoly-prl}
by substituting the classical probabilities  by the quantum ones;
i.e.,
$$
\begin{array}{lll}
p_1 &\rightarrow& q_1 (\theta ) =
{\frac{1}{2}}\left[{\mathbb I}_2 + {\bf \sigma}( \theta )\right] \otimes  {\mathbb I}_2,
\\
p_3 &\rightarrow& q_3 (\theta ) =
{\mathbb I}_2 \otimes {\frac{1}{2}}\left[{\mathbb I}_2 + {\bf \sigma}( \theta )\right],
\\
p_{ij}&\rightarrow& q_{ij} (\theta ,\theta ') =
{\frac{1}{2}}\left[{\mathbb I}_2 + {\bf \sigma}( \theta )\right]
\otimes
{\frac{1}{2}}\left[{\mathbb I}_2 + {\bf \sigma}( \theta ')\right],
\end{array}
\label{2004-qbounds-e2}
$$
with
$
{\bf \sigma}( \theta )=
\left(
\begin{array}{ll}
 \cos \theta & \sin \theta  \\
  \sin\theta & -\cos \theta
  \end{array}
\right)
$,
where $\theta $ is the relative measurement angle in the $x$--$z$-plane,
and the two particles propagate along the $y$-axis, as depicted in Fig.~\ref{2006-ql-2}.

The quantum transformation associated with the Clauser-Horne inequality for the 2--2 case
is given by
$$
\begin{array}{lcl}
O_{22}(\alpha,\beta,\gamma,\delta)
&=&  q_{13}(\alpha,\gamma) +
q_{14}(\alpha,\delta) + q_{23}(\beta,\gamma) - q_{24}(\beta,\delta)- q_{1}(\alpha) - q_{3}(\gamma)\\
&=&
{\frac{1}{2}}\left[{\mathbb I}_2 + {\bf \sigma}( \alpha )\right]
\otimes
{\frac{1}{2}}\left[{\mathbb I}_2 + {\bf \sigma}( \gamma )\right] +
{\frac{1}{2}}\left[{\mathbb I}_2 + {\bf \sigma}( \alpha )\right]
\otimes
{\frac{1}{2}}\left[{\mathbb I}_2 + {\bf \sigma}( \delta )\right] \\
&& +
{\frac{1}{2}}\left[{\mathbb I}_2 + {\bf \sigma}( \beta )\right]
\otimes
{\frac{1}{2}}\left[{\mathbb I}_2 + {\bf \sigma}( \gamma )\right]
- {\frac{1}{2}}\left[{\mathbb I}_2 + {\bf \sigma}( \beta)\right]
\otimes
{\frac{1}{2}}\left[{\mathbb I}_2 + {\bf \sigma}( \delta)\right]\\
&& -
{\frac{1}{2}}\left[{\mathbb I}_2 + {\bf \sigma}( \alpha )\right] \otimes  {\mathbb I}_2
- {\mathbb I}_2 \otimes {\frac{1}{2}}\left[{\mathbb I}_2 + {\bf \sigma}( \gamma )\right],
\end{array}
\label{2004-qbounds-e4}
$$
where $\alpha$, $\beta$, $\gamma$, $\delta$ denote the measurement angles
lying in the $x$--$z$-plane: $\alpha$ and $\beta$ for one particle, $\gamma$ and
$\delta$ for the other one.
The eigenvalues are
$$
  \label{eq:2004-qbounds-evo22}
  \lambda_{1,2,3,4}(\alpha,\beta,\gamma,\delta )
=
\frac{1}{2}\big(\pm\sqrt{1\pm\sin(\alpha -\beta )\sin(\gamma -\delta )}-1\big)
$$
yielding the maximum bound
$\|O_{22} \|= \max_{i=1,2,3,4} \lambda_i$.
Note that for the particular choice of parameters
$\alpha=0,\beta=2\theta,\gamma=\theta,\delta=3\theta$ adopted in
\cite{cabello-2003a,filipp-svo-04-qpoly}, one obtains $|O_{22}|=
\frac{1}{2}\left\{\left[ \left(3-\cos 4\theta \right)/2\right]^{1/2}
  -1\right\}\le \frac{1}{2}\left(\sqrt{2}-1\right)$, as compared to the classically allowed bound from above $0$.

\subsection{Interpretations}

The nonexistence of two-valued states on the set of quantum propositions (of greater than two-dimensional Hilbert spaces)
interpretable as truth assignments poses a great challenge for the interpretation of quantum logical propositions, relations and operations,
as well as for quantum mechanics in general.
At stake is the meaning and physical co-existence of observables which are not co-measurable.
Several interpretations have been proposed, among them contextuality, as well as the abandonment of classical omniscience and realism discussed below.

\subsubsection{Contextuality}

{\em Contextuality}
\index{contextuality}
abandons the context independence of measurement outcomes \cite{bell-66,hey-red,redhead} by supposing that
it is wrong to assume (cf. Ref.~\cite{bell-66}, Sec.~5) that the
result of an observation is independent of what observables are measured alongside of it.
Bell \cite[Sec.~5]{bell-66} states
that the
\begin{quote}
{\em ``$\ldots$
result of an observation may reasonably depend
not only on the state of the system  $\ldots$
but also on the complete disposition  of the apparatus.''}
\end{quote}
Note also Bohr's remarks \cite{bohr-1949} about
\begin{quote}
{\em ``the impossibility of any sharp separation
between the behavior of atomic objects and the interaction with the measuring instruments which serve to define
the conditions under which the phenomena appear.''}
\end{quote}

Contextuality might be criticized as an attempt to maintain omniscience and omni-realism
even in view of a lack of consistently assignable truth values on quantum propositions.
Omniscience or omni-realism is the belief that
\begin{quote}
{\em ``all observables exist even without being experienced by any finite mind.''}
\end{quote}
Contextuality supposes that an
\begin{quote}
{\em ``observable exists  without being experienced by any finite mind,
but it may have different values, depending on its context.''}
\end{quote}

So far, despite some claims to have measured contextuality,
there is no direct experimental evidence.
Some experimental findings inspired by Bell-type inequalities
\cite{aspect-81,aspect-82a,wjswz-98},
the Kochen-Specker theorem
\cite{simon-2002,hasegawa:230401}
as well as the Greenberger-Horne-Zeilinger theorem
\cite{panbdwz}
measure incompatible contexts one after another;
i.e., temporally sequentially, and not simultaneously.
Hence, different contexts can only be measured on different particles.
A more direct test of contextuality might be an EPR configuration of two quanta in three-dimensional Hilbert space
interlinked in a single observable, as discussed above.

\subsubsection{Abandonment of classical omniscience}

As has been pointed out already, contextuality might be criticized for its presumption of quantum omniscience;
in particular the supposition that
a physical system, at least in principle, is capable of ``carrying'' all answers to any
classically retrievable question.
This is true classically, since the classical context is the entirety of observables.
But it need not be true for other types of (finite) systems or agents.
Take for example, a refrigerator.
If it is automated in a way to tell you whether or not there is enough milk in it,
it will be at a complete loss at answering a totally different question, such as if there is enough oil in
the engine of your car. It is a matter of everday experience that not all agents are prepared to give answers to all
perceivable questions.

Nevertheless, if one forces an agent  to answer a question it is incapable to answer,
the agent might throw some sort of ``fair coin'' --- if it is capable of doing so --- and present random answers.
This scenario of a context mismatch between preparation and measurement is the basis of quantum random number generators
\cite{Quantis} which serve as a kind of ``quantum random oracle'' \cite{Cris04,calude-dinneen05}.
It should be kept in mind that randomness, at least algorithmically \cite{chaitin2,chaitin3,calude:02}, does not come ``for free,''
thus exhibiting an amazing capacity of single quanta to support random outcomes.
Alternatively, the unpredictable, erratic outcomes might, in the context translation \cite{svozil-2003-garda} scenario,
be due to some stochasticity originating from the interaction with
a ``macroscopic'' measurement apparatus, and the undefined.

One interpretation of the impossibility to operationalize more than a single context is the abandonment of classical omniscience:
in this view,
whereas it might be meaningful theoretically and formally to study the entirety of the context superstructure,
only a single context operationally exists.
Note that, in a similar way as {\em retrieving} information from a quantized system,
the only information {\em codable} into a quantized system is given by a single block or context.
If the block contains $n$ atoms corresponding to $n$ possible measurement outcomes, then
the information content is a $n$it \cite{zeil-99,DonSvo01,svozil-2002-statepart-prl}.
The information needs not be ``located'' at a particular particle, as it can be ``distributed'' over a multi--partite state.
In this sense,
the quantum system could be viewed as a kind of (possibly nonlocal) {\em programmable integrated circuit,}
such as a {\em field programmable qate array} or an {\em application specific integrated circuit.}

Quantum observables make only sense when interpreted as a function of some context, formalized by either some Boolean subalgebra or by the maximal operator.
It is useless in this framework to believe in the existence of a single isolated observable
devoid of the context from which it is derived.
In this holistic approach, isolated observables separated from its missing contexts do not exist.

Likwise, it is wrong to assume that
all observables which could in principle (``potentially'') have been measured, also co-exist,
irrespective of whether or not they have or could have been actually measured.
Realism in the sense of
\begin{quote}
{\em ``co-measurable entities sometimes exist without being experienced by any finite mind''}
might still be assumed for a {\em single} context, in particular the one in which the system was prepared.
\end{quote}

\subsubsection{Subjective idealism}

Still another option is subjective idealism, denying the ``existence'' of
observables which could in principle (``potentially'') have been measured,
but actually have not been  measured:
in this view, it is wrong to assume that \cite{stace}
\begin{quote}
{\em ``entities sometimes exist without being experienced by any finite mind.''}
\end{quote}
Indeed, Bekeley states \cite{berkeley},
\begin{quote}
{\em ``For as to what is said of the absolute existence of unthinking things without any relation to their being perceived,
that seems perfectly unintelligible.
Their esse [[to be]] is percepi [[to be perceived]], nor is it possible they should have any existence out of the minds or thinking things which perceive them.''}
\end{quote}
With this assumption, the Bell, Kochen-Specker and Greenberger-Horne-Zeilinger theorems and similar have merely theoretical, formal relevance for physics, because
they operate with unobservable physical ``observables'' and entities or with counterfactuals which are inferred rather than measured.

\section{Automata and generalized urn logic}

The following quasi--classical logics take up the notion of contexts as blocks representing Boolean subalgebras
and the pastings among them.
They are quasi--classical, because unlike quantum logics
they possess sufficiently many two-valued states to allow embeddings into Boolean algebras.

\subsection{Partition logic}
\index{partition logic}

The empirical logics
(i.e., the propositional calculi) associated with the
generalized urn models  suggested by
Ron Wright \cite{wright:pent,wright}, and automaton  logics
(APL)
\cite{svozil-93,schaller-96,dvur-pul-svo,cal-sv-yu,svozil-ql}
are equivalent (cf. Refs.~\cite[p.145]{svozil-ql} and \cite{svozil-2001-eua})
and can be subsumed by partition logics.
The logical equivalence of automaton models
with generalized urn models suggests that these logics are more general and ``robust''
with respect to changes of the particular model than could have been expected from
the particular instances of their first appearance.

Again the concept of context or block is very important here.
Partition logics are formed by pasting together contexts or blocks based on the {\em partitions of a set of states.}
The contexts themselves are derived from the input/output analysis of experiments.

\subsection{Generalized urn models}
\index{generalized urn model}
A generalized urn model
${\cal U}=\langle U,C,L,\Lambda \rangle $ is
characterized as follows.
Consider an ensemble of balls with black background color.
Printed on these balls are some color symbols from a symbolic alphabet $L$.
The colors are elements of a set of colors $C$.
A particular ball type is associated with a unique combination of mono-spectrally
(no mixture of wavelength) colored symbols
printed on the black ball background.
Let $U$ be the set of ball types.
We shall assume that every ball contains
just one single symbol per color.
(Not all types of balls; i.e., not all color/symbol combinations, may be present in
the ensemble, though.)

Let
$\vert U\vert $ be the number of different types of balls,
$\vert C\vert $ be the number of different mono-spectral colors,
$\vert L\vert $ be the number of different output symbols.

Consider the deterministic ``output'' or ``lookup''
function $\Lambda (u,c)=v$,
$u\in U$,
$c\in C$,
$v\in L$,
which returns one symbol per ball type and color.
One interpretation of this lookup function $\Lambda$ is as follows.
Consider a set of $\vert C\vert $ eyeglasses build from filters for the
$\vert C\vert $ different colors.
Let us assume that these mono-spectral filters are
``perfect'' in that they totally absorb light of all other colors
but a particular single one.
In that way, every color can be associated with a particular eyeglass and vice versa.

When a spectator looks at a particular ball through such an eyeglass,
the only operationally recognizable symbol will be the one in the particular
color which is transmitted through the eyeglass.
All other colors are absorbed, and the symbols printed in them will appear black
and therefore cannot be differentiated from the black background.
Hence the ball appears to carry a different ``message'' or symbol,
depending on the color at which it is viewed.
This kind of ``complementarity'' has been used for a demonstration of quantum cryptography
\cite{svozil-2005-ln1e}.

An empirical logic can be constructed as follows.
Consider the set of all ball types.
With respect to a particular colored eyeglass, this set disjointly ``decays''
or gets partitioned into those ball types which can be separated by the particular color of
the eyeglass.
Every such partition of ball types can then be identified with a Boolean algebra whose atoms are the elements of the partition.
A pasting of all of these Boolean algebras yields the empirical logic associated with
the particular urn model.

Consider, for the sake of demonstration, a single color and its associated partition of the set of ball types
(ball types within a given element of the partition cannot be differetiated by that color).
In the generalized urn model, an element $a$ of this partition is a set of ball types which corresponds to an elementary proposition
\begin{quote}
{\em ``the ball drawn from the urn is of the type contained in $a$.''}
\end{quote}

\subsection{Automaton models}
\index{automaton model}
A (Mealy type) automaton
${\cal A}=\langle S,I,O,\delta ,\lambda \rangle$ is characterized
by the set of states $S$,
by the set of input symbols $I$,
and by the set of output symbols $O$.
$\delta (s,i)=s'$ and
$\lambda (s,i)=o$,
$s,s'\in S$,
$i\in I$
and $o\in O$
represent the transition and the output functions, respectively.
The restriction to Mealy automata is for convenience only.

In the analysis of a {\em state identification problem}, a typical automaton experiment aims at
\index{state identification problem}
an operational determination of an {\em unknown initial state}
by the input of some symbolic sequence and the observation of the resulting output symbols.
Every such input/output experiment results
in a state partition in the following way.
Consider a particular automaton.
Every experiment on such an automaton which tries to
solve the initial state problem is characterized
by a set of input/output symbols
as a result of the possible input/output sequences for this experiment.
Every such distinct set of input/output symbols is associated with a set of initial automaton
states which would reproduce that sequence.
This state set may contain one or more states, depending on the ability of the experiment
to separate different initial automaton states.
A partitioning of the automaton states is obtained if one considers
a single input sequence and the variety of all
possible output sequences (given a particular automaton).
Stated differently:
given a set of inputs, the set of initial automaton states ``break down'' into disjoint
subsets associated with the possible output sequences.
(All elements of a subset yield the same output on the same input.)

This partition can then be identified with a Boolean algebra,
with the elements of the partition interpreted as atoms.
By pasting the Boolean algebras of the ``finest'' partitions together one obtains
an empirical partition logic associated with the particular automaton.
(The converse construction is also possible, but not unique; see below.)

For the sake of simplicity,
we shall assume that every experiment just deals with a single input/output
combination.
That is, the finest partitions are reached already after the first symbol.
This does not impose any restriction on the partition logic, since
given any particular automaton, it is always possible
to construct another automaton with exactly the same partition logic as the first one
with the above property.

More explicitly, given any partition logic, it is always possible to
construct a corresponding automaton with the following specification:
associate with every element of the set of partitions a single input symbol.
Then take the partition with the highest number of elements and associate a single output
symbol with any element of this partition.
(There are then sufficient output
symbols available for the other partitions as well.)
Different partitions require different input symbols;
one input symbol per partition.
The output function  can then be defined by associating a single output symbol per element
of the partition (associated with a particular input symbol).
Finally, choose a transition function which completely looses the state information
after only one transition; i.e., a transition function which maps all automaton state into
a single one.

A typical proposition in the automaton model refers to a partition element $a$  containing automaton states which cannot be distinguished
by the analysis of the strings of input and output symbols; i.e., it can be expressed by
\begin{quote}
{\em ``the automaton is initially in a state which is contained in $a$.''}
\end{quote}

\subsection{Contexts}

In the generalized urn model represent everything that is knowable by looking in only a single color.
For automata, this is equivalent to considering only a single string of input symbols.
Formally, this amounts to the identification of blocks with contexts, as in the quantum case.

\subsection{Proof of logical equivalence of automata and generalized urn models}

From the definitions and constructions mentioned in the previous sections
it is intuitively clear
that, with respect to the empirical logics,
generalized urn models and finite automata models
are equivalent.
Every logic associated with a generalized urn model
can be interpreted as an automaton partition logic
associated with some (Mealy) automaton (actually an infinity thereof).
Conversely, any logic associated with some (Mealy) automaton
can be interpreted as a logic associated with some generalized urn model
(an infinity thereof).
We shall proof these claims by explicit construction.
Essentially, the lookup function
$\Lambda$ and the output function $\lambda$ will be identified.
Again, the restriction to Mealy automata is for convenience only.
The considerations are robust with respect to variations of finite input/output automata.

\subsubsection{Direct construction of  automaton models from  generalized urn models}

In order to define an APL associated with a Mealy automaton
${\cal A}=\langle S,I,O,\delta ,\lambda \rangle$
from a generalized urn model ${\cal U}=\langle U,C,L,\Lambda \rangle $,
let
$u\in U$,
$c\in C$,
$v\in L$,
and
$s,s'\in S$,
$i\in I$,
$o\in O$, and assume
$\vert U\vert =\vert S\vert$,
$\vert C\vert =\vert I\vert$,
$\vert L\vert =\vert O\vert$.
The following identifications can be made
with the help of the bijections $t_S,t_I$ and $t_O$:
$$
\begin{array}{llllll}
t_S(u)=s, \;
 t_I(c)=i, \;
 t_O(v)=o, \\
\delta (s,i)= s_i  \quad {\rm for \; fixed\;}s_i\in S {\rm \;and \;
arbitrary\;}s\in S,\; i\in I,\\
\lambda  (s,i) = t_O\left(\Lambda (t_S^{-1}(s),t_I^{-1}(i))\right).
\end{array}
\label{oto-ua}
$$
More generally, one could use equivalence classes instead of a bijection.
Since the input-output behavior is equivalent and the
automaton transition function is trivially $\vert L\vert $-to-one,
both entities yield the same propositional calculus.

\subsubsection{Direct construction of generalized urn models from automaton models}

Conversely,
consider an arbitrary Mealy automaton ${\cal A}=\langle S,I,O,\delta ,\lambda \rangle$
and its associated propositional calculus APL.

Just as before, associate with every single automaton state $s\in S$
a ball type $u$,
associate with every input symbol $i\in I$
a unique color $c$,
and
associate with every output symbol $o\in O$
a unique symbol $v$; i.e., again
$\vert U\vert =\vert S\vert$,
$\vert C\vert =\vert I\vert$,
$\vert L\vert =\vert O\vert$.
The following identifications can be made
with the help of the bijections $\tau_U,\tau_C$ and $\tau_L$:
\begin{eqnarray*}
\begin{array}{llllll}
\tau_U(s)=u, \; \tau_C(i)=c, \;  \tau_L(o)=v, \;
\Lambda  (u,c) = \tau_L (\lambda (\tau_U^{-1}(u),\tau_C^{-1}(c))).
\end{array}
\end{eqnarray*}
A comparison yields
\begin{eqnarray*}
\begin{array}{llllll}
\tau_U^{-1}=t_S, \; \tau_C^{-1}=t_I, \;  \tau_L^{-1}=t_O.
\end{array}
\end{eqnarray*}

\subsubsection{Schemes using dispersion-free states}

Another equivalence scheme  uses the fact that both
automaton partition logics and the logic of generalized urn models
have a separating (indeed, full) set of
dispersion-free states.
Stated differently, given a finite atomic logic with a separating set of states, then
the enumeration of the complete set of dispersion-free states enables
the explicit construction of  generalized urn models and automaton logics
whose logic corresponds to the original one.

This can be achieved by ``inverting'' the set of two-valued states as follows.
(The method is probably best understood by considering the examples below.)
Let us start with an atomic logic with a separating set of states.
\begin{itemize}
\item[(i)]
In the first step, every atom of this lattice is labeled by some natural number,
starting from ``$1$'' to ``$n$'', where $n$  stands for the number of lattice atoms.
The set of atoms is denoted by $A=\{1,2,\ldots , n\}$.

\item[(ii)]
Then, all two-valued states of this lattice are labeled consecutively
by natural numbers, starting from ``$m_1$'' to ``$m_r$'', where $r$  stands for the number of
two-valued states.
The set of states is denoted by $M=\{m_1,m_2,\ldots , m_r\}$.

\item[(iii)]
Now  partitions are defined as follows.
For every atom, a set is created whose members are the numbers or ``labels'' of
the two-valued states which are ``true'' or take on the value ``1'' on this atom.
More precisely,
the elements $p_i(a)$ of the partition ${\cal P}_j$ corresponding to
some atom $a\in A$ are defined by
$$p_i (a) =
\left\{
k \mid m_k(a)=1, \; k\in M
\right\}
.
$$
The partitions are obtained by taking the unions of all $p_i$ which belong to the same
subalgebra ${\cal P}_j$.
That the corresponding sets are indeed partitions follows from the properties of
two-valued states: two-valued states (are ``true'' or) take on the value ``$1$'' on just one atom
per subalgebra and (``false'' or) take on the value ``$0$'' on all other atoms of this subalgebra.

\item[(iv)]
Let there be $t$ partitions labeled by ``1'' through ``$t$''.
The partition logic is obtained by a pasting of all partitions
${\cal P}_j$, $1\le j \le t$.

\item[(v)] In the following step, a corresponding generalized urn model or automaton model is
obtained from the partition logic just constructed.

\begin{itemize}
\item[(a)] A generalized urn model is obtained by the following identifications
(see also \cite[p. 271]{wright:pent}).
\begin{itemize}
\item[$\bullet$]
Take as many ball types as there are two-valued states; i.e.,
$r$ types of balls.
\item[$\bullet$]
Take as many colors as there are subalgebras or partitions; i.e., $t$ colors.
\item[$\bullet$]
Take as many symbols as there are elements in the partition(s) with the maximal number of elements;
i.e., $\max_{1\le j\le t}\vert {\cal P}_j\vert \le n$.
To make the construction easier, we may just take as many symbols as there are atoms; i.e., $n$ symbols.
(In some cases, much less symbols will suffice).
Label the symbols by $v_l$.
Finally, take $r$ ``generic'' balls with black background.
Now associate with every measure a different ball type.
(There are $r$ two-valued states, so there will be $r$ ball types.)
\item[$\bullet$]
The $i$th ball type is painted by colored
symbols  as follows:
Find the atoms  for which the $i$th two-valued state $m_i$ is $1$.
Then paint the symbol corresponding to every such lattice atom on the ball, thereby choosing
the color associated with the subalgebra or partition
the atom belongs to.
If the atom belongs to more than one subalgebra,
then paint the same symbol in as many colors as there are partitions or subalgebras
the atom belongs to (one symbol per subalgebra).
\end{itemize}
This completes the construction.

\item[(b)] A Mealy automaton is obtained by the following identifications
(see also \cite[pp. 154--155]{svozil-93}).
\begin{itemize}
\item[$\bullet$]
Take as many automaton states as there are two-valued states; i.e.,
$r$  automaton states.
\item[$\bullet$]
Take as many input symbols as there are subalgebras or partitions; i.e., $t$ symbols.
\item[$\bullet$]
Take as many output symbols as there are elements in the partition(s) with the maximal number of elements
(plus one additional auxiliary output symbol ``$\ast$'', see below);
i.e., $\max_{1\le j\le t}\vert {\cal P}_j\vert \le n+1$.
\item[$\bullet$]
The output function is chosen to match the elements of the state partition corresponding
to some input symbol.
Alternatively, let the lattice atom $a_q\in A$
must be an atom of the subalgebra corresponding to the input $i_l$.
Then one may choose an output function such as
$$
\lambda (m_k,i_l)= \left\{
\begin{array}{l}
a_q \quad{\rm if }\;m_k (a_q)= 1\\
\ast \; \quad {\rm if }\;m_k (a_q)= 0\\
\end{array}
\right.
$$
with
$1\le k \le r$
and
$1\le l \le t$.
Here, the additional output symbol ``$\ast$'' is needed.

\item[$\bullet$]
The transition function is $r$--to--1 (e.g., by $\delta (s,i)=s_1$, $s,s_1\in S$,
$i\in I$), i.e., after one input the information about the
initial state is completely lost.
\end{itemize}
This completes the construction.
\end{itemize}
\end{itemize}

\subsubsection{Example 1: The automaton partition logic $L_{12}$}

In what follows we shall illustrate the above constructions with a couple of examples.
First, consider the generalized urn model
$$\langle \{u_1,\ldots ,u_5\},\{{\rm \color{red}red,\color{green}green}\},\{1,\ldots ,5\},\Lambda \rangle$$
with $\Lambda$ listed in Table \ref{t-wright}(a).
\begin{table}
{\footnotesize
\begin{center}
\begin{tabular}{cc}
\begin{tabular}{|c|cc|}
\hline\hline
ball type & \color{red}red & \color{green}green \\ \hline
1  & \color{red}$1$ & \color{green}$3$ \\
2  & \color{red}$1$ & \color{green}$4$ \\
3  & \color{red}$2$ & \color{green}$3$ \\
4  & \color{red}$2$ & \color{green}$4$ \\
5  & \color{red}$5$ & \color{green}$5$ \\
\hline\hline
\end{tabular}
\qquad
\qquad
&
\begin{tabular}{|c|ccccc|ccccc|}
 \hline\hline
 &&&$\delta$ && & &&$\lambda$&&\\
\raisebox{2.5ex}[0cm][0cm]{state} &1&2&3&4&5 & 1&2&3&4&5\\
 \hline
0&1&1&1&1&1 & 1&1&2&2&5\\
1&1&1&1&1&1 & 3&4&3&4&5\\
 \hline\hline
\end{tabular}
\\
$\;$\\
(a)&(b)\\
\end{tabular}
\end{center}
}
\caption{\label{t-wright} (a) Ball types in Wright's generalized urn model
\protect\cite{wright} (cf. also \protect\cite[p.143ff]{svozil-ql}).
(b) Transition and output table of an associated automaton model.}
\end{table}

The associated Mealy automaton can be directly constructed as follows.
Take $t_S=t_O= {\rm id}$, where $ {\rm id}$ represents the identity function,
and take
$t_I({\color{red}\rm red})=0$
and
$t_I({\color{green}\rm green})=1$,
respectively.
Furthermore, fix a (five$\times$two)-to-one transition function by $\delta(.,.)=1$.
The transition and output tables are listed in Table  \ref{t-wright}(b).
Both empirical structures yield the same propositional logic $L_{12}$
which is depicted in Fig.~\ref{2004-qnc-f1}(b).

\subsubsection{Example 2: The generalized urn logic $L_{12}$}
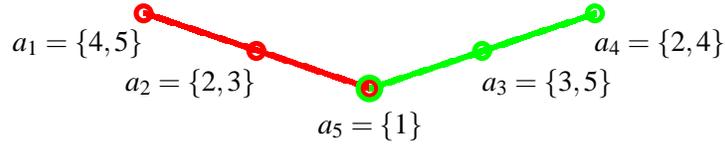
\begin{figure}
\begin{center}
\unitlength 1.00mm
\allinethickness{2pt}
\begin{picture}(60.16,15.17)
\multiput(0.16,15.17)(0.36,-0.12){84}{\color{red}\line(1,0){0.36}}
\multiput(30.16,5.17)(0.36,0.12){84}{\color{green}\line(1,0){0.36}}
\put(30.16,5.17){\color{red}\circle{2.00}}
\put(30.16,5.17){\color{green}\circle{3.00}}
\put(45.16,10.17){\color{green}\circle{2.00}}
\put(60.16,15.17){\color{green}\circle{2.00}}
\put(0.16,15.17){\color{red}\circle{2.00}}
\put(15.16,10.17){\color{red}\circle{2.00}}
\put(60.16,11.17){\makebox(0,0)[lc]{$a_4=\{ 2,4\}$}}
\put(45.16,5.84){\makebox(0,0)[lc]{$a_3=\{ 3,5\}$}}
\put(30.33,0.17){\makebox(0,0)[cc]{$a_5=\{ 1\}$}}
\put(15.16,5.84){\makebox(0,0)[rc]{$a_2=\{ 2,3\}$}}
\put(0.16,11.17){\makebox(0,0)[rc]{$a_1=\{ 4,5\}$}}
\end{picture}
\end{center}
\caption{\label{2008-ql-f-pcac} Atoms are identified with the set of the numbers of
two-valued measures which are $1$ on that atom.}
\end{figure}

Let us start with
an automaton whose transition and output tables are listed in Table~\ref{t-wright}(b)
and indirectly construct a logically equivalent generalized urn model by using dispersion-free states.
The first thing to do is to figure out all dispersion-free states of $L_{12}$ depicted in Fig.~\ref{2008-ql-f-pcac}
[see also Fig.~\ref{2004-qnc-f1}(b)].
There are five of them, which might be written in vector form; i.e., in lexicographic order:
$$
\begin{array}{ccccccccccccc}
      &a_1,&a_2,&a_3,&a_4,&a_5\\
m_1= (&0,&0,&0,&0,&1&),   \\
m_1= (&0,&0,&0,&0,&1&),   \\
m_2= (&0,&1,&0,&1,&0&),   \\
m_3= (&0,&1,&1,&0,&0&),   \\
m_4= (&1,&0,&0,&1,&0&),  \\
m_5= (&1,&0,&1,&0,&0&).
\end{array}
$$

Now define the following generalized urn model as follows.
Note that the associated logic contains two subalgebras with the atoms
$a_1,a_2,a_5$ and $a_3,a_4,a_5$, respectively,
which are interlinked at $a_5$.
There are five two-valued measures corresponding to five ball types.
They are colored according to the coloring rules listed in Table
\ref{t-wright1}.
\begin{table}
{\footnotesize
\begin{center}
\begin{tabular}{|c|ccccc|ccccc|}
\hline\hline
&\multicolumn{10}{c|}{colors}\\
& \multicolumn{5}{c}{$c_1$}  & \multicolumn{5}{c|}{$c_2$}   \\
\raisebox{2.5ex}[0cm][0cm]{ball type} & \multicolumn{5}{c}{\color{red}``red''}& \multicolumn{5}{c|}{\color{green}``green''}   \\ \hline
1  & $\ast $& $\ast $ &$\ast $&$\ast$& \color{red}5      & $\ast $&$\ast$& $\ast $& $\ast$ & \color{green}5 \\
2  & $\ast $& \color{red}2       &$\ast $&$\ast$& $\ast$ & $\ast $&$\ast$& $\ast $& \color{green}4      & $\ast$ \\
3  & $\ast $& \color{red}2       &$\ast $&$\ast$& $\ast$ & $\ast $&$\ast$& \color{green}3      & $\ast $& $\ast$ \\
4  & \color{red}1      & $ \ast $&$\ast $&$\ast$& $\ast$ & $\ast $&$\ast$& $\ast$ & \color{green}4      & $\ast$ \\
5  & \color{red}1      & $ \ast $&$\ast $&$\ast$& $\ast$ & $\ast $&$\ast$& \color{green}3      & $\ast $& $\ast$ \\
\hline\hline
\end{tabular}
\end{center}
}
\caption{\label{t-wright1} Representation of the sign coloring scheme $\Lambda$.
``$\ast$'' means no sign at all (black) for the corresponding atom.}
\end{table}

For every single atom of the logic,
one could just write down the {\em set of numbers} of the two-valued measures
(in some enumeration of all two-valued measures) which are $1$ on that atom.
For $L_{12}$, with the contexts defined by $\{a_1,a_2,a_5\}$ and $\{a_3,a_4,a_5\}$ and the measures defined above,
this construction is depicted in Fig.~\ref{2008-ql-f-pcac}.
Then,  one chooses a single color per block, as well as the symbols $1,2,3$ indicating the atom.
Now, generate as many ball types as there are blocks,
and paint one of the symbols ``$1$,'' ``$2$'' or ``$3$'' in the color
associated with the elements of the partitions of the two-valued measures.

\subsubsection{Example 3: generalized urn model of the Kochen-Specker ``bug'' logic}
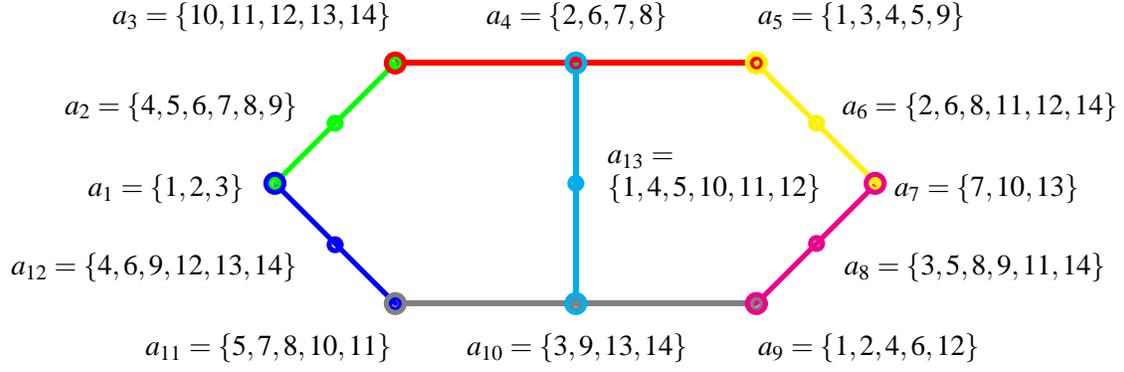
\begin{figure}
\begin{center}
\unitlength 0.8mm
\allinethickness{2pt}
\begin{picture}(108.00,55.00)
\put(25.00,7.33){\color{gray}\line(1,0){60.00}}
\put(25.00,47.33){\color{red}\line(1,0){60.00}}
\put(55.00,7.33){\color{cyan}\line(0,1){40.00}}
\put(25.00,7.33){\color{blue}\line(-1,1){20.00}}
\put(5.00,27.33){\color{green}\line(1,1){20.00}}
\put(85.00,7.33){\color{magenta}\line(1,1){20.00}}
\put(105.00,27.33){\color{yellow}\line(-1,1){20.00}}
\put(24.67,55.00){\makebox(0,0)[rc]{$a_3=\{10,11,12,13,14\}$}}
\put(55.33,55.00){\makebox(0,0)[cc]{$a_4=\{2,6,7,8\}$}}
\put(85.33,55.00){\makebox(0,0)[lc]{$a_5=\{1,3,4,5,9\}$}}
\put(9.00,40.00){\makebox(0,0)[rc]{$a_2=\{4,5,6,7,8,9\}$}}
\put(99.33,40.00){\makebox(0,0)[lc]{$a_6=\{2,6,8,11,12,14\}$}}
\put(0.00,26.33){\makebox(0,0)[rc]{$a_1=\{1,2,3\}$}}
\put(108.00,26.33){\makebox(0,0)[lc]{$a_7=\{7,10,13\}$}}
\put(60.33,31.33){\makebox(0,0)[lc]{$a_{13}=$}}
\put(60.33,26.33){\makebox(0,0)[lc]{$\{1,4,5,10,11,12\}$}}
\put(9.00,13.33){\makebox(0,0)[rc]{$a_{12}=\{4,6,9,12,13,14\}$}}
\put(99.67,13.33){\makebox(0,0)[lc]{$a_8=\{3,5,8,9,11,14\}$}}
\put(24.67,-0.05){\makebox(0,0)[rc]{$a_{11}=\{5,7,8,10,11\}$}}
\put(55.33,-0.05){\makebox(0,0)[cc]{$a_{10}=\{3,9,13,14\}$}}
\put(85.33,-0.05){\makebox(0,0)[lc]{$a_9=\{1,2,4,6,12\}$}}
\put(15.00,17.09){\color{blue}\circle{2.00}}
\put(25.00,7.33){\color{blue}\circle{2.00}}
\put(25.00,7.33){\color{gray}\circle{3.00}}
\put(55.00,27.33){\color{cyan}\circle{2.00}}
\put(85.00,7.33){\color{gray}\circle{2.00}}
\put(85.00,7.33){\color{magenta}\circle{3.00}}
\put(95.00,17.33){\color{magenta}\circle{2.00}}
\put(5.00,27.33){\color{green}\circle{2.00}}
\put(5.00,27.33){\color{blue}\circle{3.0}}
\put(15.00,37.33){\color{green}\circle{2.00}}
\put(25.00,47.33){\color{green}\circle{2.00}}
\put(25.00,47.33){\color{red}\circle{3.00}}
\put(55.00,47.33){\color{red}\circle{2.00}}
\put(55.00,47.33){\color{cyan}\circle{3.00}}
\put(85.00,47.33){\color{red}\circle{2.00}}
\put(85.00,47.33){\color{yellow}\circle{3.00}}
\put(55.00,7.33){\color{gray}\circle{2.00}}
\put(55.00,7.33){\color{cyan}\circle{3.00}}
\put(104.76,27.33){\color{yellow}\circle{2.00}}
\put(104.76,27.33){\color{magenta}\circle{3.00}}
\put(95.00,37.33){\color{yellow}\circle{2.00}}
\end{picture}
\end{center}
\caption{\label{2001-cesena-f2} Greechie diagram of automaton partition logic
with a nonfull set of dispersion-free measures.}
\end{figure}

Another, less simple example, is a logic which is already mentioned by Kochen
and Specker \cite{kochen1} (this is a subgraph of their $\Gamma_1$)
whose automaton partition logic is depicted in Fig. \ref{2001-cesena-f2}.
(It is called ``bug'' by Professor Specker  \cite{Specker-priv} because of the similar shape with a bug.)
There are 14 dispersion-free states which are listed in Table \ref{2001-cesena-t2}(a).
The associated generalized urn model is listed in Table \ref{2001-cesena-t2}(b).
\begin{table}
{\footnotesize
\begin{center}
{ {\footnotesize
\setlength{\tabcolsep}{3pt}
\begin{tabular}{|c|ccccccccccccc||ccccccc|}
\hline\hline
&\multicolumn{13}{c||}{(a) lattice atoms}&\multicolumn{7}{|c|}{(b) colors}\\
\raisebox{1.5ex}[0cm][0cm]{$m_r$ and}&$a_1$&$a_2$&$a_3$&$a_4$&$a_5$&$a_6$&$a_7$&$a_8$&$a_9$&$a_{10}$&$a_{11}$&$a_{12}$&$a_{13}$&\color{green}$c_1$&\color{red}$c_2$&\color{yellow}$c_3$&\color{magenta}$c_4$&$\color{gray}c_5$&$\color{blue}c_6$&$\color{cyan}c_7$\\
\raisebox{1.5ex}[0cm][0cm]{ball type}&&&&&&&&&&&&&&&&&&&&\\
\hline
1  &1&0&0&0&1&0&0&0&1&0&0&0&1&  \color{green}1&\color{red}1&\color{yellow} 1&\color{magenta} 1&\color{gray} 1&\color{blue} 1&\color{cyan}1          \\
2  &1&0&0&1&0&1&0&0&1&0&0&0&0&  \color{green}1&\color{red}2&\color{yellow} 2&\color{magenta} 1&\color{gray} 1&\color{blue} 1&\color{cyan}2           \\
3  &1&0&0&0&1&0&0&1&0&1&0&0&0&  \color{green}1&\color{red}1&\color{yellow} 1&\color{magenta} 2&\color{gray} 2&\color{blue} 1&\color{cyan}3          \\
4  &0&1&0&0&1&0&0&0&1&0&0&1&1&  \color{green}2&\color{red}1&\color{yellow} 1&\color{magenta} 1&\color{gray} 1&\color{blue} 2&\color{cyan}1          \\
5  &0&1&0&0&1&0&0&1&0&0&1&0&1&  \color{green}2&\color{red}1&\color{yellow} 1&\color{magenta} 2&\color{gray} 3&\color{blue} 3&\color{cyan}1          \\
6  &0&1&0&1&0&1&0&0&1&0&0&1&0&  \color{green}2&\color{red}2&\color{yellow} 2&\color{magenta} 1&\color{gray} 1&\color{blue} 2&\color{cyan}2           \\
7  &0&1&0&1&0&0&1&0&0&0&1&0&0&  \color{green}2&\color{red}2&\color{yellow} 3&\color{magenta} 3&\color{gray} 3&\color{blue} 3&\color{cyan}2           \\
8  &0&1&0&1&0&1&0&1&0&0&1&0&0&  \color{green}2&\color{red}2&\color{yellow} 2&\color{magenta} 2&\color{gray} 3&\color{blue} 3&\color{cyan}2           \\
9  &0&1&0&0&1&0&0&1&0&1&0&1&0&  \color{green}2&\color{red}1&\color{yellow} 1&\color{magenta} 2&\color{gray} 2&\color{blue} 2&\color{cyan}3          \\
10 &0&0&1&0&0&0&1&0&0&0&1&0&1&  \color{green}3&\color{red}3&\color{yellow} 3&\color{magenta} 3&\color{gray} 3&\color{blue} 3&\color{cyan}1          \\
11 &0&0&1&0&0&1&0&1&0&0&1&0&1&  \color{green}3&\color{red}3&\color{yellow} 2&\color{magenta} 2&\color{gray} 3&\color{blue} 3&\color{cyan}1          \\
12 &0&0&1&0&0&1&0&0&1&0&0&1&1&  \color{green}3&\color{red}3&\color{yellow} 2&\color{magenta} 1&\color{gray} 1&\color{blue} 2&\color{cyan}1          \\
13 &0&0&1&0&0&0&1&0&0&1&0&1&0&  \color{green}3&\color{red}3&\color{yellow} 3&\color{magenta} 3&\color{gray} 2&\color{blue} 2&\color{cyan}3          \\
14 &0&0&1&0&0&1&0&1&0&1&0&1&0&  \color{green}3&\color{red}3&\color{yellow} 2&\color{magenta} 2&\color{gray} 2&\color{blue} 2&\color{cyan}3          \\
\hline\hline
\end{tabular}
} }
\end{center}
}
\caption{(a) Dispersion-free states  of the Kochen-Specker ``bug'' logic with 14 dispersion-free states
and (b) the associated generalized urn model (all blank entries ``$\ast$''have been omitted).
\label{2001-cesena-t2}}
\end{table}

\subsection{Probability theory}

The probability theory of partition logics is based on a full set of state, allowing to define probabilities via the convex sum
of those states. This is essentially the same procedure as for classical probabilities.
In the same way, bounds on probabilities can be found through the computation of the faces of correlation polytopes.

Consider, as an example, a logic already discussed.
Its automaton partition logic is depicted in Fig. \ref{2001-cesena-f2}.
The  correlation polytope of this lattice consists of 14 vertices listed
in Table \ref{2001-cesena-t2}, where  the 14 rows indicate the vertices
corresponding to the 14 dispersion-free states. The columns
represent the partitioning of the automaton states.
The solution of the hull problem  yields the equalities
$$
\label{2001-cesena-eq18}
\begin{array}{l}
1 =   P_1  +P_2  +P_3 =   P_4 +P_{10} +P_{13}                          ,\\
1 = P_1+P_2 -P_4  +P_6+P_7 =  -P_2+P_4 -P_6+P_8 -P_{10}+P_{12}                  ,\\
1 =   P_1  +P_2 -P_4+P_6-P_8+P_{10}  +P_{11} ,\\
0 = P_1  +P_2-P_4 -P_5  = -P_1-P_2  +P_4 -P_6  +P_8  +P_9.
\end{array}
$$
The operational meaning of $P_i=P_{a_i}$  is
``the probability to find the automaton in state $a_i$.''
The above equations are equivalent to all probabilistic conditions on the contexts
(subalgebras)
$1
=P_1+P_2+P_3
=P_3+P_4+P_5
=P_5+P_6+P_7
=P_7+P_8+P_9
=P_9+P_{10}+P_{11}
=P_4+P_{10}+P_{13}
$.

Let us now turn to the joint probability case.
Notice that formally it is possible to form a statement such as $a_1\wedge a_{13}$
(which would be true for measure number $1$ and false otherwise),
but this is not operational on a single automaton,
since no experiment can decide such a proposition on a single automaton.
Nevertheless, if one considers a ``singlet state'' of two automata which are in an unknown
yet identical initial state, then an expression such as $a_1\wedge a_{13}$ makes operational sense
if property $a_1$ is measured on the first automaton
and property $a_{13}$ on the second automaton. Indeed, all joint probabilities
$a_i\wedge a_j\wedge \ldots a_n$
make sense  for $n$-automaton singlets.

\section{Summary}

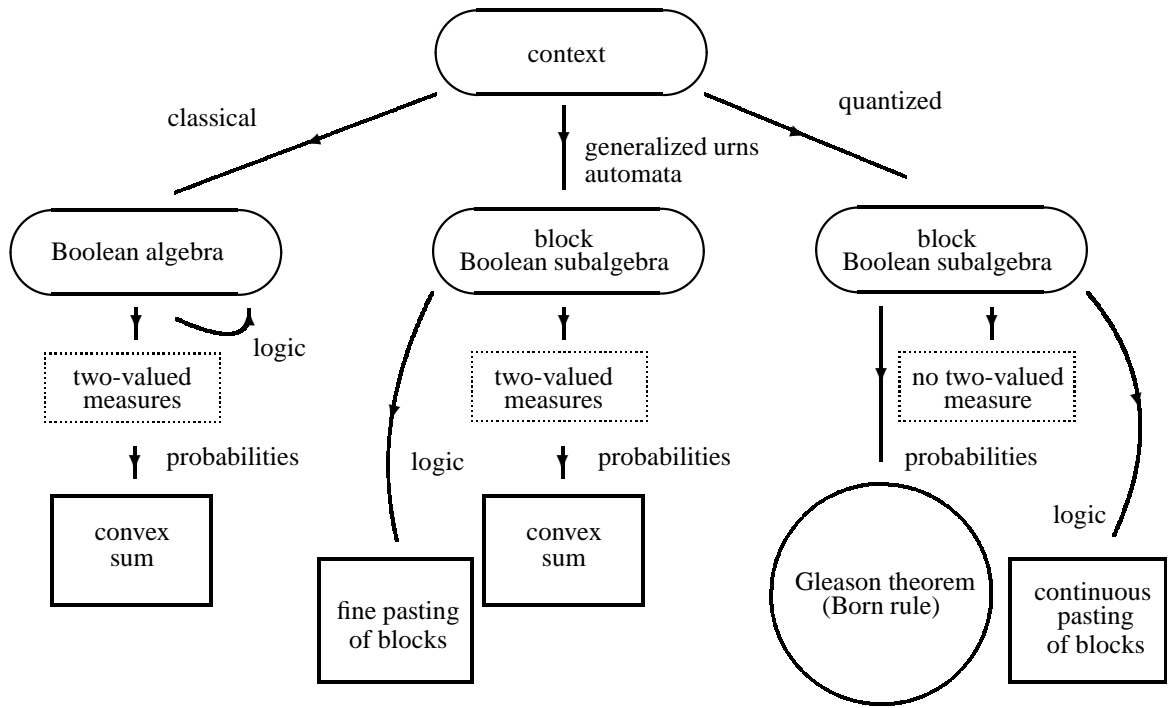
\begin{figure}
\begin{center}
{\footnotesize
\unitlength .7mm 
\allinethickness{1pt} 
\ifx\plotpoint\undefined\newsavebox{\plotpoint}\fi 
\begin{picture}(303.75,133.25)(0,0)
\put(106.125,125.25){\oval(51.25,16)[]}
\put(25.375,87.25){\oval(51.25,16)[]}
\put(105.625,87.5){\oval(51.25,16)[]}
\put(178.375,87.75){\oval(51.25,16)[]}
\put(105.5,125){\makebox(0,0)[cc]{context}}
\put(105,90.25){\makebox(0,0)[cc]{block}}
\put(105,85.25){\makebox(0,0)[cc]{Boolean subalgebra}}
\put(24,87.25){\makebox(0,0)[cc]{Boolean algebra}}
\put(177.75,90.25){\makebox(0,0)[cc]{block}}
\put(177.75,85.25){\makebox(0,0)[cc]{Boolean subalgebra}}
\put(80.25,117.5){\makebox(0,0)[cc]{}}
\put(55.875,108.125){\vector(-3,-1){.128}}\multiput(80.75,117.5)(-.1625816993,-.0612745098){306}{\line(-1,0){.1625816993}}
\put(105,107.375){\vector(0,-1){.128}}\put(105,115.25){\line(0,-1){15.75}}
\put(150.875,109.625){\vector(3,-1){.128}}\multiput(131.75,117.5)(.1488326848,-.0612840467){257}{\line(1,0){.1488326848}}
\put(6.5,55.25){\makebox(33.25,13)[cc]{}}
\multiput(6.372,68.122)(0,-.92857){15}{{\rule{.4pt}{.4pt}}}
\multiput(6.372,55.122)(.977941,0){35}{{\rule{.4pt}{.4pt}}}
\multiput(6.372,68.122)(.977941,0){35}{{\rule{.4pt}{.4pt}}}
\multiput(39.622,68.122)(0,-.92857){15}{{\rule{.4pt}{.4pt}}}
\put(23,64){\makebox(0,0)[cc]{two-valued}}
\put(23,59){\makebox(0,0)[cc]{measures}}
\put(87.5,55.5){\makebox(33.25,13)[cc]{}}
\multiput(87.372,68.372)(0,-.92857){15}{{\rule{.4pt}{.4pt}}}
\multiput(87.372,55.372)(.977941,0){35}{{\rule{.4pt}{.4pt}}}
\multiput(87.372,68.372)(.977941,0){35}{{\rule{.4pt}{.4pt}}}
\multiput(120.622,68.372)(0,-.92857){15}{{\rule{.4pt}{.4pt}}}
\put(103,64){\makebox(0,0)[cc]{two-valued}}
\put(103,59){\makebox(0,0)[cc]{measures}}
\put(168.75,55.5){\makebox(33.25,13)[cc]{}}
\multiput(168.622,68.372)(0,-.92857){15}{{\rule{.4pt}{.4pt}}}
\multiput(168.622,55.372)(.977941,0){35}{{\rule{.4pt}{.4pt}}}
\multiput(168.622,68.372)(.977941,0){35}{{\rule{.4pt}{.4pt}}}
\multiput(201.872,68.372)(0,-.92857){15}{{\rule{.4pt}{.4pt}}}
\put(185.5,64){\makebox(0,0)[cc]{no two-valued}}
\put(185.5,59){\makebox(0,0)[cc]{measure}}
\put(46.75,113.25){\makebox(0,0)[rc]{classical}}
\put(109,107.25){\makebox(0,0)[lc]{generalized urns}}
\put(109,102.25){\makebox(0,0)[lc]{automata}}
\put(157.25,116){\makebox(0,0)[lc]{quantized}}
\put(165.25,62.5){\vector(0,-1){.128}}\put(165.25,77.25){\line(0,-1){29.5}}
\put(186.185,22.5){\line(0,1){.9537}}
\put(186.163,23.454){\line(0,1){.9517}}
\put(186.098,24.405){\line(0,1){.9478}}
\multiput(185.989,25.353)(-.05051,.31396){3}{\line(0,1){.31396}}
\multiput(185.838,26.295)(-.04857,.2335){4}{\line(0,1){.2335}}
\multiput(185.643,27.229)(-.05916,.23104){4}{\line(0,1){.23104}}
\multiput(185.407,28.153)(-.055696,.182485){5}{\line(0,1){.182485}}
\multiput(185.128,29.066)(-.053293,.149799){6}{\line(0,1){.149799}}
\multiput(184.808,29.965)(-.060062,.147215){6}{\line(0,1){.147215}}
\multiput(184.448,30.848)(-.057177,.123708){7}{\line(0,1){.123708}}
\multiput(184.048,31.714)(-.05491,.105853){8}{\line(0,1){.105853}}
\multiput(183.609,32.561)(-.059675,.103241){8}{\line(0,1){.103241}}
\multiput(183.131,33.387)(-.05717,.089258){9}{\line(0,1){.089258}}
\multiput(182.617,34.19)(-.061177,.086561){9}{\line(0,1){.086561}}
\multiput(182.066,34.969)(-.058552,.075316){10}{\line(0,1){.075316}}
\multiput(181.481,35.722)(-.056293,.065973){11}{\line(0,1){.065973}}
\multiput(180.861,36.448)(-.05924,.06334){11}{\line(0,1){.06334}}
\multiput(180.21,37.144)(-.062064,.060575){11}{\line(-1,0){.062064}}
\multiput(179.527,37.811)(-.064759,.057685){11}{\line(-1,0){.064759}}
\multiput(178.815,38.445)(-.074052,.060142){10}{\line(-1,0){.074052}}
\multiput(178.074,39.047)(-.076715,.056706){10}{\line(-1,0){.076715}}
\multiput(177.307,39.614)(-.088021,.059058){9}{\line(-1,0){.088021}}
\multiput(176.515,40.145)(-.09062,.054986){9}{\line(-1,0){.09062}}
\multiput(175.699,40.64)(-.10466,.057151){8}{\line(-1,0){.10466}}
\multiput(174.862,41.097)(-.122463,.059798){7}{\line(-1,0){.122463}}
\multiput(174.005,41.516)(-.12506,.054157){7}{\line(-1,0){.12506}}
\multiput(173.129,41.895)(-.14863,.056471){6}{\line(-1,0){.14863}}
\multiput(172.237,42.234)(-.181258,.059569){5}{\line(-1,0){.181258}}
\multiput(171.331,42.532)(-.183784,.051249){5}{\line(-1,0){.183784}}
\multiput(170.412,42.788)(-.23241,.05353){4}{\line(-1,0){.23241}}
\multiput(169.483,43.002)(-.31281,.05718){3}{\line(-1,0){.31281}}
\multiput(168.544,43.174)(-.31509,.04287){3}{\line(-1,0){.31509}}
\put(167.599,43.302){\line(-1,0){.9501}}
\put(166.649,43.388){\line(-1,0){.9531}}
\put(165.696,43.43){\line(-1,0){.954}}
\put(164.742,43.428){\line(-1,0){.9529}}
\put(163.789,43.383){\line(-1,0){.9499}}
\multiput(162.839,43.295)(-.31496,-.04381){3}{\line(-1,0){.31496}}
\multiput(161.894,43.164)(-.31264,-.05811){3}{\line(-1,0){.31264}}
\multiput(160.956,42.989)(-.23225,-.05422){4}{\line(-1,0){.23225}}
\multiput(160.027,42.773)(-.18363,-.051798){5}{\line(-1,0){.18363}}
\multiput(159.109,42.514)(-.181079,-.06011){5}{\line(-1,0){.181079}}
\multiput(158.204,42.213)(-.148461,-.056914){6}{\line(-1,0){.148461}}
\multiput(157.313,41.872)(-.124897,-.05453){7}{\line(-1,0){.124897}}
\multiput(156.439,41.49)(-.122284,-.060164){7}{\line(-1,0){.122284}}
\multiput(155.583,41.069)(-.104489,-.057463){8}{\line(-1,0){.104489}}
\multiput(154.747,40.609)(-.090455,-.055257){9}{\line(-1,0){.090455}}
\multiput(153.933,40.112)(-.087844,-.05932){9}{\line(-1,0){.087844}}
\multiput(153.142,39.578)(-.076545,-.056935){10}{\line(-1,0){.076545}}
\multiput(152.377,39.008)(-.073872,-.060363){10}{\line(-1,0){.073872}}
\multiput(151.638,38.405)(-.064587,-.057878){11}{\line(-1,0){.064587}}
\multiput(150.927,37.768)(-.061883,-.06076){11}{\line(-1,0){.061883}}
\multiput(150.247,37.1)(-.05905,-.063516){11}{\line(0,-1){.063516}}
\multiput(149.597,36.401)(-.056096,-.06614){11}{\line(0,-1){.06614}}
\multiput(148.98,35.674)(-.058326,-.07549){10}{\line(0,-1){.07549}}
\multiput(148.397,34.919)(-.060919,-.086743){9}{\line(0,-1){.086743}}
\multiput(147.849,34.138)(-.056904,-.089429){9}{\line(0,-1){.089429}}
\multiput(147.336,33.333)(-.059367,-.103419){8}{\line(0,-1){.103419}}
\multiput(146.861,32.506)(-.054593,-.106016){8}{\line(0,-1){.106016}}
\multiput(146.425,31.658)(-.056808,-.123878){7}{\line(0,-1){.123878}}
\multiput(146.027,30.79)(-.059623,-.147394){6}{\line(0,-1){.147394}}
\multiput(145.669,29.906)(-.052846,-.149957){6}{\line(0,-1){.149957}}
\multiput(145.352,29.006)(-.055151,-.182651){5}{\line(0,-1){.182651}}
\multiput(145.076,28.093)(-.05847,-.23122){4}{\line(0,-1){.23122}}
\multiput(144.843,27.168)(-.04787,-.23364){4}{\line(0,-1){.23364}}
\multiput(144.651,26.234)(-.04957,-.31411){3}{\line(0,-1){.31411}}
\put(144.502,25.291){\line(0,-1){.9481}}
\put(144.397,24.343){\line(0,-1){.9519}}
\put(144.334,23.391){\line(0,-1){2.859}}
\put(144.408,20.532){\line(0,-1){.9475}}
\multiput(144.519,19.585)(.05144,-.3138){3}{\line(0,-1){.3138}}
\multiput(144.674,18.643)(.04927,-.23335){4}{\line(0,-1){.23335}}
\multiput(144.871,17.71)(.05985,-.23086){4}{\line(0,-1){.23086}}
\multiput(145.11,16.787)(.056241,-.182318){5}{\line(0,-1){.182318}}
\multiput(145.391,15.875)(.05374,-.149639){6}{\line(0,-1){.149639}}
\multiput(145.714,14.977)(.060502,-.147035){6}{\line(0,-1){.147035}}
\multiput(146.077,14.095)(.057546,-.123537){7}{\line(0,-1){.123537}}
\multiput(146.48,13.23)(.055225,-.105689){8}{\line(0,-1){.105689}}
\multiput(146.922,12.385)(.059983,-.103063){8}{\line(0,-1){.103063}}
\multiput(147.401,11.56)(.057436,-.089087){9}{\line(0,-1){.089087}}
\multiput(147.918,10.758)(.055292,-.07774){10}{\line(0,-1){.07774}}
\multiput(148.471,9.981)(.058776,-.075141){10}{\line(0,-1){.075141}}
\multiput(149.059,9.23)(.056489,-.065804){11}{\line(0,-1){.065804}}
\multiput(149.68,8.506)(.059429,-.063163){11}{\line(0,-1){.063163}}
\multiput(150.334,7.811)(.062244,-.06039){11}{\line(1,0){.062244}}
\multiput(151.019,7.147)(.064931,-.057491){11}{\line(1,0){.064931}}
\multiput(151.733,6.514)(.074231,-.059921){10}{\line(1,0){.074231}}
\multiput(152.475,5.915)(.076884,-.056477){10}{\line(1,0){.076884}}
\multiput(153.244,5.35)(.088197,-.058795){9}{\line(1,0){.088197}}
\multiput(154.038,4.821)(.090784,-.054716){9}{\line(1,0){.090784}}
\multiput(154.855,4.329)(.10483,-.056838){8}{\line(1,0){.10483}}
\multiput(155.694,3.874)(.122641,-.059433){7}{\line(1,0){.122641}}
\multiput(156.552,3.458)(.125221,-.053784){7}{\line(1,0){.125221}}
\multiput(157.429,3.081)(.148798,-.056027){6}{\line(1,0){.148798}}
\multiput(158.321,2.745)(.181435,-.059028){5}{\line(1,0){.181435}}
\multiput(159.229,2.45)(.183936,-.050701){5}{\line(1,0){.183936}}
\multiput(160.148,2.197)(.23257,-.05284){4}{\line(1,0){.23257}}
\multiput(161.079,1.985)(.31298,-.05625){3}{\line(1,0){.31298}}
\multiput(162.018,1.817)(.31522,-.04193){3}{\line(1,0){.31522}}
\put(162.963,1.691){\line(1,0){.9504}}
\put(163.914,1.608){\line(1,0){.9532}}
\put(164.867,1.569){\line(1,0){.954}}
\put(165.821,1.573){\line(1,0){.9528}}
\put(166.773,1.621){\line(1,0){.9496}}
\multiput(167.723,1.712)(.31483,.04475){3}{\line(1,0){.31483}}
\multiput(168.668,1.846)(.31246,.05905){3}{\line(1,0){.31246}}
\multiput(169.605,2.023)(.23209,.05492){4}{\line(1,0){.23209}}
\multiput(170.533,2.243)(.183474,.052346){5}{\line(1,0){.183474}}
\multiput(171.451,2.505)(.180899,.06065){5}{\line(1,0){.180899}}
\multiput(172.355,2.808)(.14829,.057357){6}{\line(1,0){.14829}}
\multiput(173.245,3.152)(.124734,.054903){7}{\line(1,0){.124734}}
\multiput(174.118,3.537)(.122103,.060528){7}{\line(1,0){.122103}}
\multiput(174.973,3.96)(.104317,.057775){8}{\line(1,0){.104317}}
\multiput(175.807,4.422)(.09029,.055526){9}{\line(1,0){.09029}}
\multiput(176.62,4.922)(.087667,.059582){9}{\line(1,0){.087667}}
\multiput(177.409,5.458)(.076375,.057163){10}{\line(1,0){.076375}}
\multiput(178.173,6.03)(.073692,.060583){10}{\line(1,0){.073692}}
\multiput(178.91,6.636)(.064414,.05807){11}{\line(1,0){.064414}}
\multiput(179.618,7.275)(.061701,.060945){11}{\line(1,0){.061701}}
\multiput(180.297,7.945)(.058861,.063692){11}{\line(0,1){.063692}}
\multiput(180.944,8.646)(.055898,.066308){11}{\line(0,1){.066308}}
\multiput(181.559,9.375)(.058101,.075664){10}{\line(0,1){.075664}}
\multiput(182.14,10.132)(.060659,.086925){9}{\line(0,1){.086925}}
\multiput(182.686,10.914)(.056636,.089598){9}{\line(0,1){.089598}}
\multiput(183.196,11.72)(.059058,.103596){8}{\line(0,1){.103596}}
\multiput(183.668,12.549)(.054277,.106179){8}{\line(0,1){.106179}}
\multiput(184.103,13.399)(.056438,.124047){7}{\line(0,1){.124047}}
\multiput(184.498,14.267)(.059182,.147571){6}{\line(0,1){.147571}}
\multiput(184.853,15.152)(.052398,.150114){6}{\line(0,1){.150114}}
\multiput(185.167,16.053)(.054606,.182815){5}{\line(0,1){.182815}}
\multiput(185.44,16.967)(.05778,.23139){4}{\line(0,1){.23139}}
\multiput(185.671,17.893)(.04717,.23378){4}{\line(0,1){.23378}}
\multiput(185.86,18.828)(.04863,.31425){3}{\line(0,1){.31425}}
\put(186.006,19.771){\line(0,1){.9484}}
\put(186.109,20.719){\line(0,1){1.781}}
\put(166,25){\makebox(0,0)[cc]{Gleason theorem}}
\put(165.25,20){\makebox(0,0)[cc]{(Born rule)}}
\put(90,20.75){\framebox(29.75,20.25)[]{}}
\put(8,20.75){\framebox(29.75,20.25)[]{}}
\put(105,34){\makebox(0,0)[cc]{convex}}
\put(105,29){\makebox(0,0)[cc]{sum}}
\put(23,34){\makebox(0,0)[cc]{convex}}
\put(23,29){\makebox(0,0)[cc]{sum}}
\put(111.5,48.25){\makebox(0,0)[lc]{probabilities}}
\put(29.5,48.25){\makebox(0,0)[lc]{probabilities}}
\put(169.75,48.25){\makebox(0,0)[lc]{probabilities}}
\put(58.75,6){\framebox(28.5,23)[cc]{}}
\put(73.5,19){\makebox(0,0)[cc]{fine pasting}}
\put(73.5,14){\makebox(0,0)[cc]{of blocks}}
\put(190,6){\framebox(29,23)[cc]{}}
\put(206,23){\makebox(0,0)[cc]{continuous }}
\put(206,18){\makebox(0,0)[cc]{pasting}}
\put(206,13){\makebox(0,0)[cc]{of blocks}}
\put(72.25,56){\vector(-1,-4){.128}}\qbezier(80,79.75)(67.875,55.75)(73.25,32.75)
\put(214.25,58.25){\vector(1,-4){.128}}\qbezier(204.5,79.75)(221.5,59.75)(209.5,33.75)
\put(45.25,76.5){\vector(0,1){.128}}\qbezier(31.5,74.75)(45.625,68.375)(45.25,76.5)
\put(76,47.25){\makebox(0,0)[lc]{logic}}
\put(208,37.25){\makebox(0,0)[rc]{logic}}
\put(56,69){\makebox(0,0)[rc]{logic}}
\put(44.5,70.75){\makebox(0,0)[lc]{}}
\put(44.25,71.75){\makebox(0,0)[rc]{}}
\put(46.25,71){\makebox(0,0)[lc]{}}
\put(23.75,74){\vector(0,-1){1.5}}\put(23.75,77){\line(0,-1){6}}
\put(23.5,47.5){\vector(0,-1){1.5}}\put(23.5,50.5){\line(0,-1){6}}
\put(105,74){\vector(0,-1){1.5}}\put(105,77){\line(0,-1){6}}
\put(104.75,47.5){\vector(0,-1){1.5}}\put(104.75,50.5){\line(0,-1){6}}
\put(186.25,74){\vector(0,-1){1.5}}\put(186.25,77){\line(0,-1){6}}
\end{picture}
}
\end{center}
\caption{\label{2008-ql} ``Mind map'' representing the use of contexts to build up logics and construct probabilities.}
\end{figure}

Regarding contexts; i.e., the maximum collection of co-measurable observables, three different cases have been discussed.
The first, classical case, is characterized by omniscience.
Within the classical framework, all observables form a {\em single} context,
and everything that is in principle knowable is also knowable simultaneously.
Classical probability can be based upon the convex combinations of all two-valued states.
Fig.~\ref{2008-ql} depicts a ``mind map'' representing the use of contexts to build up logics and construct probabilities.

In the generalized urn or automaton cases, if one sticks to the rules ---
that is, if one does not view the object unfiltered
or ``screw the automaton box open'' ---
omniscience is impossible and a quasi--classical sort of complementarity emerges: depending on the color (or input string) chosen,
one obtains knowledge of a particular observable or context.
All other contexts are hidden to the experimenter unable to lift the bounds of one color filter
or one input sequence.
A system science issue is emerging here; namely the question of how intrinsic observers
perform inside of a given system \cite{svozil-93,svozil-94}.
The situation resembles quantum mechanics even more if reversible systems are
considered; where an experiment can be ``undone'' only by investing all the information gained from
previous experiments (without being able to copy these)\cite{greenberger2,hkwz}.
All incompatible blocks or contexts are pasted together to form the partition logic.
These pasting still allow a sufficient number of two-valued states for the construction of probabilities  based upon the convex combinations thereof.

In the quantum case, the Hilbert lattices can formally be thought of as pastings of a continuum of blocks or contexts,
but the mere assumption of the physical existence --- albeit inaccessible to
an intrisic observer --- of even a finite number of contexts yields a complete contradiction.
In view of this, one can adopt at least two interpretations: that
an observable depends on its context; or
that more than one context for quantum systems has no operational meaning.
The former view has been mentioned by Bell (and also by Bohr to some degree),
and can be subsumed by the term ``contextuality.''
To the author, contextuality is the last resort of a realism
which is inclined to maintain ``a sort of'' classical omniscience, even in view
of the Kochen-Specker and Bell-type theorems.

The latter viewpoint --- that quantum systems do not encode more than a single context ---
abandons omniscience, but needs to cope with the fact that it {\em is} indeed possible to
measure different contexts; even if there is a mismatch
between the preparation and the measurement context.
It has been proposed that in these cases the measurement apparatus ``translates'' one
context into the other at the prize of randomizing the measurement result \cite{svozil-2003-garda}.
This {\em context translation principle} could be tested by
changing the measurement apparatus' ability of translation.

All in all, contexts seem to be an exciting subject. The notion may become more useful and relevant,
as progress is made towards a better comprehension of the quantum world
and its differences with respect to other classical and quasi--classical systems.


\end{document}